\documentclass[sigconf]{acmart}
\AtBeginDocument{%
  \providecommand\BibTeX{{%
    \normalfont B\kern-0.5em{\scshape i\kern-0.25em b}\kern-0.8em\TeX}}}

\copyrightyear{2020}
\acmYear{2020}
\setcopyright{acmlicensed}\acmConference[SIGIR '20]{Proceedings of the 43rd
International ACM SIGIR Conference on Research and Development in
Information Retrieval}{July 25--30, 2020}{Virtual Event, China}
\acmBooktitle{Proceedings of the 43rd International ACM SIGIR Conference on
Research and Development in Information Retrieval (SIGIR '20), July 25--30,
2020, Virtual Event, China}
\acmPrice{15.00}
\acmDOI{10.1145/3397271.3401167}
\acmISBN{978-1-4503-8016-4/20/07}
\acmSubmissionID{270}

\newcommand{\ie}{\emph{i.e., }}

\newcommand{\wrt}{\emph{w.r.t. }}

\usepackage{enumitem}
\usepackage{verbatim}
\usepackage[ruled]{algorithm2e}
\usepackage{multirow}
\usepackage{subfigure} 
\usepackage{ragged2e}

\begin{document}
%\title{Efficient Recommender Retraining with \\ Sequential Meta-Learning}
\fancyhead{}
\title{How to Retrain Recommender System? A Sequential Meta-Learning Method$^*$}
% \author{Yang Zhang, Xiangnan He,Fuli Feng,Chenxu Wang,Meng Wang,Yan Li}
% \email{zy2015@mail.ustc.edu.cn,xiangnanhe,fulifeng93@gmail.com,wcx123@mail.ustc.edu.cn,}
% \email{}
\author{Yang Zhang$^1$, Fuli Feng$^2$, Chenxu Wang$^1$, Xiangnan He$^1$, Meng Wang$^3$, Yan Li$^4$, Yongdong Zhang$^1$}

\thanks{$^*$This work is supported by the National Natural Science Foundation of China (61972372, U19A2079,61725203). Fuli Feng is the Corresponding Author.}

\affiliation{\institution{$^1$University of Science and Technology of China, $^2$National University of Singapore\\
$^3$Hefei University of Technology, $^4$Beijing Kuaishou Technology Co., Ltd. Beijing, China}}

\email{{fulifeng93,xiangnanhe}@gmail.com, {zy2015,wcx123}@mail.ustc.edu.cn}
\email{eric.mengwang@gmail.com,liyan@kuaishou.com,zhyd73@ustc.edu.cn} 

\def\authors{Yang Zhang, Fuli Feng, Chenxu Wang, Xiangnan He, Meng Wang, Yan Li, Yongdong Zhang}
% \email{hongrc@hfut.edu.cn, kanmy@comp.nus.edu.sg, chuats@comp.nus.edu.sg}

%\settopmatter{printacmref=false, printfolios=false}
% \author{Yang Zhang}
% \affiliation{%
%   \institution{University of Science and Technology of China}
% }
% \email{zy2015@mail.ustc.edu.cn}
% \author{Xiangnan He}
% \authornote{corresponding author}
% \affiliation{%
%   \institution{University of Science and Technology of China}
% }
% \email{xiangnanhe@gmail.com}
% \author{Fuli Feng}
% \affiliation{%
%   \institution{National University of Singapore}
% }
% \email{fulifeng93@gmail.com}
% \author{Chenxu Wang}
% \affiliation{%
%   \institution{University of Science and Technology of China}
% }
% \email{wcx123@mail.ustc.edu.cn}
% \author{Meng Wang}
% \affiliation{%
%   \institution{Hefei University of Technology}
% }
% \email{eric.mengwang@gmail.com}
% \author{Yan Li}
% \affiliation{%
%   \institution{Kuaishou Inc.}
% }
% \email{liyan@kuaishou.com}
% \author{Yongdong Zhang}
% \affiliation{%
%   \institution{University of Science and Technology of China}
% }
% \email{zhyd73@ustc.edu.cn}

\begin{abstract}
Practical recommender systems need be periodically retrained to refresh the model with new interaction data. To pursue high model fidelity, it is usually desirable to retrain the model on both historical and new data, since it can account for both long-term and short-term user preference. However, a full model retraining could be very time-consuming and memory-costly, especially when the scale of historical data is large. In this work, we study the model retraining mechanism for recommender systems, a topic of high practical values but has been relatively little explored in the research community. 

Our first belief is that retraining the model on historical data is unnecessary, since the model has been trained on it before. Nevertheless, normal training on new data only may easily cause overfitting and forgetting issues, since the new data is of a smaller scale and contains fewer information on long-term user preference. To address this dilemma, we propose a new training method, aiming to abandon the historical data during retraining through learning to transfer the past training experience. 
Specifically, we design a neural network-based transfer component, which transforms the old model to a new model that is tailored for future recommendations. 
To learn the transfer component well, we optimize the ``future performance'' --- i.e., the recommendation accuracy evaluated in the next time period. 
Our \textit{Sequential Meta-Learning} (SML) method offers a general training paradigm that is applicable to any differentiable model. We demonstrate SML on matrix factorization and conduct experiments on two real-world datasets. 
Empirical results show that SML not only achieves significant speed-up, but also outperforms the full model retraining in recommendation accuracy, validating the effectiveness of our proposals. We release our codes at: \url{https://github.com/zyang1580/SML}.
%Codes are available upon acceptance. 

\end{abstract}





\maketitle
\section{Introduction}\label{sec:intro}
%Background, RS, retraining RS is important
\begin{figure}
    \centering
    \includegraphics[width=0.45\textwidth]{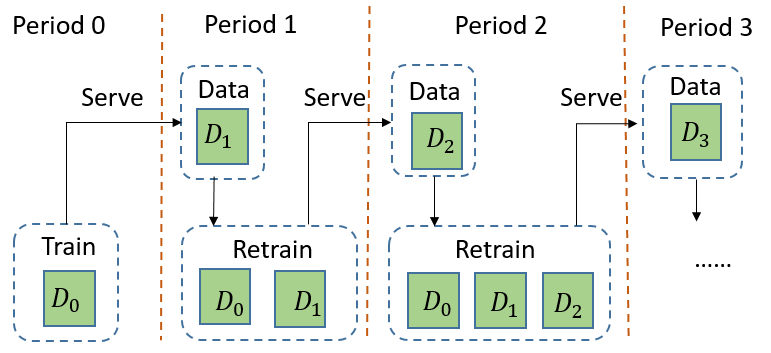}\vspace{-5pt}
    \caption{An illustration of periodical model retraining.}\vspace{-15pt}
    \label{fig:retraining}
\end{figure}

Recommender systems play an increasingly important role in the current Web 2.0 era which faces with serious information overload issues. The key technique in a recommender system is the personalization model, which estimates the preference of a user on items based on the historical user-item interactions~\cite{BPR,LightGCN}. 
Since users keep interacting with the system, new interaction data is collected continuously, providing the latest evidence on user preference. 
Therefore, it is important to retrain the model with the new interaction data, so as to provide timely personalization and avoid being stale~\cite{hidden}. 
With the increasing complexity of recommender models, it is technically challenging to apply real-time updates on the models in an online fashion, especially for those expressive but computationally expensive deep neural networks~\cite{NFM, xDeepFM, KGAT}.
As such, a common practice in industry is to perform model retraining periodically, for example, on a daily or weekly basis. 
Figure \ref{fig:retraining} illustrates the model retraining process. 

%Existing work: three types of retraining: full, only new, reservior-based; their limitations
Intuitively, the historical interactions provide more evidence on user long-term (e.g., inherent) interest and the newly collected interactions are more reflective of user short-term preference. To date, three retraining strategies are most widely adopted, depending on the data utilization:
\begin{itemize}[leftmargin=*]
    \item Fine-tuning, which updates the model based on the new interactions only~\cite{online-update,NMRN}. This way is memory and time efficient, since only new data is to be handled. However, it ignores the historical data that contains long-term preference signal, thus can easily cause overfitting and forgetting issues ~\cite{real-time}.
    \item Sample-based retraining, which samples historical interactions and adds them to new interactions to form the training data~\cite{SIGIR18-YinHongzhi,real-time}. The sampled interactions are expected to retain long-term preference signal, which need be carefully selected to obtain representative interactions. In terms of recommendation accuracy, it is usually worse than using all historical interactions due to the information loss caused by sampling~\cite{SIGIR18-YinHongzhi}. 
    \item Full retraining, which trains the model on the whole data that includes all historical and new interactions. Undoubtedly, this method costs most resources and training time, but it provides the highest model fidelity since all available interactions are utilized.
\end{itemize}

While the above three strategies have their pros and cons, we argue a key limitation is that they lack an explicit optimization towards the retraining objective --- i.e., the retrained model should serve well for the recommendations of the next time period. In practice, user interactions of the next time period provide the most important evidence on the generalization performance of the current model, and are usually used for model selection or validation.
As such, an effective retraining method should take this objective into account and formulate the retraining process towards optimizing the objective, a much more principled way than manually crafting heuristics to select data examples~\cite{SIGIR18-YinHongzhi,real-time,online-update,sampling-reservoir}. %cite 3-4 reservoir-based work. 

%Opportunities and challenges in doing retraining
In this work, we explore the central theme of model retraining in recommendation, a topic of high practical value in industry recommender systems but receives relatively little scrutiny in research. Although full model retraining provides the highest fidelity, we argue that it is not necessary to do so. The key reason is that the historical interactions have been trained in the previous training, which means the model has already distilled the ``knowledge'' from the historical data. If there is a way to retain the knowledge well and transfer it to the training on new interactions, we should be able to keep the same performance level as the full retraining, even though we do not use the historical data during model retraining. Furthermore, if the knowledge transfer is ``smart'' enough to capture more patterns like recent data is more reflective of near future performance, we even have the opportunity to improve over the full retraining in recommendation accuracy. 

%Our work
To this end, we propose a new retraining method with two major considerations: (1) building an expressive component that transfers the knowledge gained in previous training to the training on new interactions, and (2) optimizing the transfer component towards the recommendation performance in the near future.
To achieve the first goal, we devise the transfer component as a convolutional neural network (CNN), which inputs the previous model parameters as constant and the present model as trainable parameters. 
The rationality is that the knowledge gained in previous training is condensed in model parameters, such that an expressive neural network should be able to distill the knowledge towards the desired purpose. 
To achieve the second goal, in addition to normal training on newly collected interactions, we further train the transfer CNN on the future interactions of next time period. As such, the CNN can learn how to combine the old parameters with present parameters, with the objective of predicting the user interactions of the near future. 
The whole architecture can be seen as an instance of meta-learning~\cite{finn2017MAML}: the retraining of each time period is a task, which has the new interactions of the current period as the training set and the future interactions of the next period as the testing set. By learning to train historical tasks well, we expect the method to perform well for future tasks. 
Since our meta-learning mechanism is operated on sequential data, we name it as \textit{Sequential Meta-Learning} (SML). 

%Figure \ref{fig:method} illustrates our method framework.  The TRANSFER component inputs the old model as constant parameters and the present model as trainable parameters, and outputs a new model to serve for the next stage recommendations.  The output model is expected to perform well on both 
%Specifically, we divide the parameters of our method into two components: the new recommender model that is learned to perform well on new interactions, and the transfer component that \

%Summary of contributions
The main contributions of this work are summarized as follows:
\begin{itemize}[leftmargin=*]
    \item We highlight the importance of recommender retraining research and formulate the sequential retraining process as an optimizable problem.  
    \item We propose a new retraining approach that is 1) efficient by training on new interactions only, and 2) effective by optimizing for the future recommendation performance. 
    \item We conduct experiments on two real-world datasets of Adressa news and Yelp business. Extensive results demonstrate the effectiveness and rationality of our method.  %Empirical results demonstrate improvements over full training, sequential recommender models\cite{Caser,GRU4Rec}, and recent streaming recommender systems\cite{SIGIR18-YinHongzhi}, justifying the effectiveness of our method. 
\end{itemize}

%SPACE
%The remainder of the paper is organized as follows. We first present problem formulation in Section \ref{sec:problem} and review related work in Section \ref{sec:related}. We then elaborate our method in Section \ref{sec:method} and conduct experiments in Section \ref{sec:experiment}. Finally, we conclude the paper in Section \ref{sec:conclude}.  

\section{Problem Formulation}\label{sec:problem}
%In this section, we will give the formulation of the retraining problem and some symbols annotations.

In real-world recommender systems, user interaction data streams in continuously. To keep the predictive model fresh with recent data, a common choice is to retrain the model periodically. We represent the data as $\{ D_0 , \dots, D_t, D_{t+1},\dots\}$, where $D_t$ denotes the data newly collected in the time period $t$. Assume each retraining is triggered right after $D_t$ is collected. 
A period can be any length of time, e.g., daily, weekly or until a certrain number of interactions are collected, depending on the system requirement and implementation abilty. 

In the retraining of time period $t$, the system has access to all previous data, i.e., $\{D_0,\dots,D_{t-1}\}$, and the new data $D_t$. 
Since the retrained model is used to serve for the near future, it is reasonable to judge its effectiveness based on $D_{t+1}$ --- the data collected in the next time period. As such, we set the recommendation performance on $D_{t+1}$ as the generalization goal of the $t$-th period retraining. 
Let the model parameters after the $t$-th peirod retraining be $W_t$. We treat each retraining as a \textit{task}, formulating it as:
\begin{equation}\label{def:task-def}
(\{D_m: m \leq t \}, W_{t-1}) \xrightarrow{get}  W_t \xleftarrow{test} D_{t+1}.
\end{equation}
That is, based on all accessible data at the time of retraining and the model parameters of the previous retraining, we aim to get a new set of model parameters that can perform well on the near future data $D_{t+1}$. 
The mostly used solution in industry is to perform a full retraining on the whole data with $W_{t-1}$ as the initialization. This solution is straightforward to implement. However, the drawback is that it takes too many computation resources, a relative long retraining time, and requires to enlarge the computation power as time goes by. Another limitation is that the full retraining lacks explicit optimization for the performance on $D_{t+1}$. This is non-trivial to address, since directly using $D_{t+1}$ in training will cause information leak and worse generalization ability. 

%In this work, we concentrate on model retraining. In this scenario, the data is streaming coming. With time goes by, we get dataset set $\{ D_0 , \dots, D_t, D_{t+1},\dots\}$, where $D_t$ is the data collected in time period $t$. A time period can be any long time that it's demanded,such as several days,or a few weeks. In any time period t, we have the collected  data $D_t$, it is new data,which have not been used for training the recommendation model,but  may have been used as test data , \ie the new data is collected after having been tested in online scenario. Besides the new collected data, we also have a model teained with previous data contained in $\{D_0,\dots,D_{t-1}\}$, and we take the parameters of the model to represented it, and denote it as $W_{t-1}$, the subscript $t-1$ represent the model parameters obtain in time period $t-1$. With the data $\{D_{m}:(m\leq t)\}$ and the previous model $W_{t-1}$, the goal of retraining the model is to get a new model $W_t$, and we except it will perform well in the near future data $D_{t+1}$, this process can be treat as a task,and We can write it as  the following tuple form :
%\begin{equation}\label{def:task-def}
%(\{D_m: m \leq t \}, W_{t-1}) \xrightarrow{to \,\, get}  W_t \xleftarrow{test} D_{t+1}
%\end{equation}

In this work, we aim to utilize the newly collected data $D_{t}$ only plus the previous model parameters $W_{t-1}$, so as to pursue a good retrained model as evaluated on $D_{t+1}$. Thus we reformulate the retraining process as:
%if we limit only the new data ($D_t$) can be used, the (\ref{def:task-def}) can be simply written as :
\begin{equation}\label{def:simple-task}
(D_t, W_{t-1}) \xrightarrow{get}  W_t \xleftarrow{test} D_{t+1},
\end{equation}
which we denote as the task $\tau_t$. For $\tau_0$, the previous model parameters are just random initialization. 
A straightforward solution is to perform stochastic gradient descent (SGD) updates on $D_t$ with $W_{t-1}$ as initialization. However, it is easy to encounter the forgetting issue of user long-term interest, since the effect of initialization is weakening with more updates. Moreover, this solution also lacks optimization scheme towards serving $D_{t+1}$. 
%we denote (\ref{def:simple-task}) as task $\tau_t$, t is time period. 
%In the following, if without special instruction, we take care on (\ref{def:simple-task}), because we aim at only using new data and previous model to get a new model which can obtain comparable performance with full or sample-based retraining method.  

%Considering the different time, we have a sequence of task $\{\tau_{m},\dots,\tau_t,\tau_{t+1},\dots \}$, it should be noticed that the task is sequential appearing, so only if the $\tau_t$ have been completed , the $\tau_{t+1}$ can be solved. For task $\tau_t$, our goal is the good performance in the future data $D_{t+1}$. What's more, we except that after solving the task  will not cause the future task $\tau_{k}$, where $k>t$, become hard to be solved, i.e. we except  each task in the sequence tasks can perform well.

Distinct from the definition of $task$ in standard meta-learning~\cite{finn2017MAML,TAML}, the tasks here naturally form a sequence $\{\tau_0, ..., \tau_t, \tau_{t+1}, ... \}$.
In online serving (testing), only if $\tau_t$ has been completed we can move to $\tau_{t+1}$. As such, the offline training should follow the similar manner of sequential training to ensure the method can generalize well in future serving. 
Lastly, addressing the problem can be seen as an instance of meta-learning, since the learning target is how to solve the tasks well (i.e., with a good generalization ability on future tasks), which is a higher-level problem than simply learning model parameters on $D_t$.
\section{METHOD}\label{sec:method}
%We present our proposed method in this section. 
Firstly, we present the model overview to solve the task $\tau_t$, the core of which is to design a transfer component that effectively converts the old model $W_{t-1}$ to a new model $W_t$. Then, we elaborate our design of the transfer. Next, we discuss how to train the model 
with good performance on current data $D_t$ as well as good generalization to future data $D_{t+1}$. 
Lastly, we demonstrate how to instantiate our generic method on matrix factorization, one of the most classic and representative models for collaborative filtering. 

%In this section, we will first introduce the overall framework that can sovle $\tau_t$, the introduce the details of components of the framework. Then, we will introduce how to train it and apply to sequence tasks $\{\tau_0,\dots,\tau_t,\dots\}$. At the last, we give an implementation based on MF for this framework. 

\subsection{Model Overview}\label{sec:method-overall}%\vspace{-5pt}
\begin{figure} 
	\centering
	\includegraphics[width=0.4\textwidth]{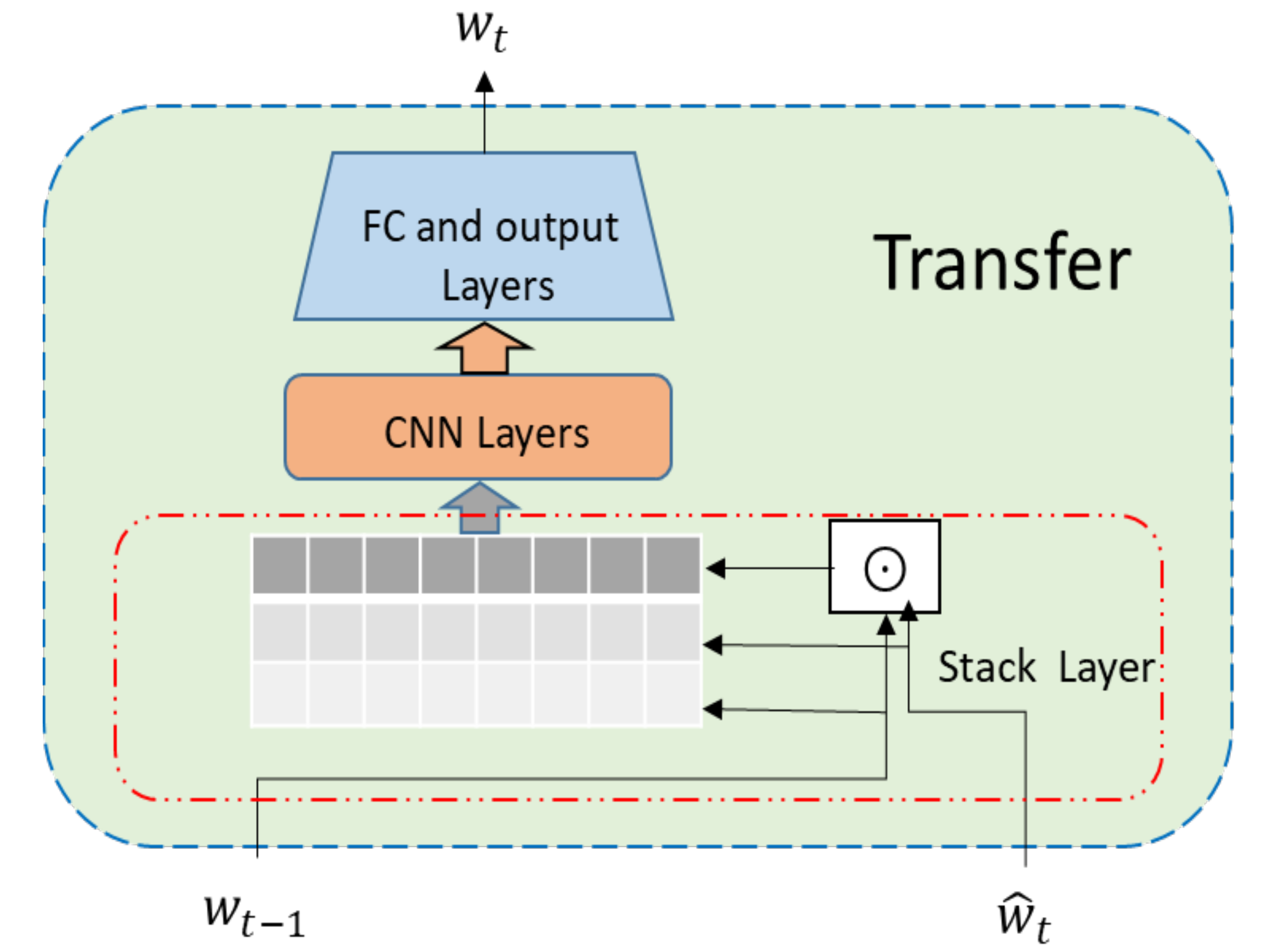}\vspace{-5pt}
	\caption{Model overview of our transfer-based retraining for the $t$-th time period. $W_{t-1}$ represents the previous recommender, $\hat{W_{t}}$ is a recommender learned on new data $D_t$ only. The transfer component is to combine the ``knowledge'' in $W_{t-1}$ and $\hat{W_{t}}$ to obtain the new recommender $W_t$ for serving the next period.} \vspace{-5pt}
	\label{fig:overall-framework}
\end{figure}

We aim to solve the task $\tau_t$ defined in Equation (\ref{def:simple-task}) which leverages only the new data $D_t$ to achieve a comparable or even better performance than the full retraining. The belief is that the past data $\{D_0, ..., D_{t-1}\}$ have been seen in previous training, such that the ``knowledge'' useful for recommendation has been gained and stored in model parameters $W_{t-1}$. 
Another consideration is to make our method technically applicable to many recommender models, rather than a specific one. 

To this end, we design a generic model framework, as illustrated in Figure~\ref{fig:overall-framework}. 
It has three components: 1) $W_{t-1}$ represents the previous recommender model that is trained from past data, 2) $\hat{W}_t$ denotes a new recommender model that needs to be learned from the current data $D_t$, and 3) Transfer is the module to combine the ``knowledge'' contained in $W_{t-1}$ and $\hat{W}_t$ to form a new recommender model $W_t$, which is used for serving next period recommendations. 
In the $t$-th period retraining, $W_{t-1}$ is set as constant input, and the retraining consists of two main steps:
%Instead of solving the task defined in \eqref{def:task-def}, we turn to solve the task $\tau_t$ defined in \eqref{def:simple-task} to achieve a comparable performance with the belief that the $W_{t-1}$ has already distilled the "knowledge" of  the historical data. To achieve this goal, we design an uniform framework can be applied to any differentiable recommendation model, which is shown in the Figure~\ref{fig:overall-framework}. It has three components, $W_{t-1}$ represent the previous rec model which gained from previous data, $\hat{W_t}$ is a rec model which need to be learned from current data $D_t$, and transfer $\Theta$ is a module used to combine the "knowledge" contained in $W_{t-1}$ and $\hat{W_t}$ to form a new rec model $W_t$. At time t, after we get the $W_{t-1}$ and $D_t$, there are two thing we need to do:
\begin{itemize}
	\item [1.] Obtaining $\hat{W_t}$, which is expected to contain  useful signal for recommendation  from $D_t$. This step can be done by optimizing standard recommendation loss,
	%like pointwise~\cite{NCF} or pairwise loss~\cite{BPR}, 
	denoted as $L_{r} (\hat{W_t} | D_t) $. 
	\item[2.] Obtaining $W_t$, which is the output of the transfer module: 
	\begin{equation}\label{equation:overall-TRfunction} 
	W_t = f_{\Theta}(W_{t-1}, \hat{W_t})
	\end{equation}
	where $f_{\Theta}$ denotes the transfer function, $\Theta$ denotes its parameters, and 
	$W_{t-1}$ and $\hat{W}_t$ are its input.
\end{itemize}

In this framework, $W_{t-1}$ and $\hat{W_t}$ can be any differentiable recommender model, as long as they are of the same architecture (i.e., the parameter number and semantics are the same). Only the transfer component needs to be carefully designed, which is our contribution to be introduced next. 

% are two rec model entities, not different from the original rec model. So the transfer component is the only part need to be designed. We will introduce it in the following.

%%%%%%%%%%%%%%%%%%%%%%%%%%%%%%%%%%%%%%%%%%%%%%%%%%%%%%%%%%%%%%%%%%%%%%%%%%%%%%%%%%%%%%%%%%%%%%%%%%%%%%%%%%%%%%%%%%%%%%%%%%%%%%%%%%%%%%%%%
%            transfer
%
%
\subsection{Transfer Design}
% \begin{figure} 
%     \centering
%     \includegraphics[width=0.5\textwidth]{picture/MF_transferpng.jpg}
%     \caption{\textbf{cnn-based transfer}}. 
%     \Description{training process }
%     \label{fig:MF-based-transfer}
% \end{figure}
Functionally speaking, the transfer combines  parameters $W_{t-1}$ and $\hat{W_t}$ to form a new group of parameters $W_{t}$. 
As the most basic requirement, $W_t$ needs be of the same shape with $W_{t-1}$ and $\hat{W_t}$. 
This requirement can be easily satisfied by operations like weighted sum:
$$W_t = \alpha W_{t-1}  + (1-\alpha)\hat{W_t},$$
where $\alpha$ is the combination coefficient which can be either pre-defined or learned. The method is simple to interpret by paying different attentions to previous and current trained knowledge; it is also easy to train, since few parameters are introduced. 
However it has limited representation ability, for example, cannot account for the relations between different dimensions of parameters.  

For expressiveness of the transfer, multi-layer perceptron (MLP) can be another option: 
$$W_t = MLP(W_{t-1}||\hat{W_t}).$$
Despite the universal approximation theorem of MLP~\cite{mlpapproximation}, it may be practically difficult to be trained well~\cite{NFM,latent-cross}. Another limitation is that it does not emphasize the interactions beweeen the parameters of the same dimension, which could be important for understanding parameter evolution. 
As an example, suppose the model is matrix factorization and the parameters are user embedding. 
Then the difference $\hat{W_t} - W_{t-1}$ means parameter change which can capture the interest drift; and each dimension of the product $W_{t-1} \odot \hat{W_t}$ indicates the importance of the dimension in reflecting user interest of both short-term and long-term. However, MLP lacks mechanisms to explicitly capture such patterns. 

%As an example, suppose the parameters be user embedding. 
 
%And the second, for $w_u$ and $w_i$, they are corresponding parameters in two model, obviously, the values of  $w_{t-1}$ and $\hat{w_t}$ in the same dimension have strong correlation, and it will tell us some important information. Such as the difference $\Delta = \hat{w_t} - w_{t-1}$ can tell us the trend of parameters changing and the mean $w_m = \frac{1}{2}(w_{t-1}+\hat{w_t})$ tell us it importance level (such as, for user embedding in CF model, the changing one dimension value of a user embedding will tell us the user one interest drifts, the mean of it can tell us the degree of interest importance). 

To this end, we design the transfer component to be capable of not only emphasizing the relation between $W_{t-1}$ and $\hat{W_t}$ at each dimension, but also capturing the relations among different dimensions. 
Inspired by the success of convolutional neural network (CNN) in capturing local-region features in image processing, we design the transfer based on CNN. The CNN architecture can be found in the green box of Figure \ref{fig:overall-framework}, which consists of a stack layer, two convolution layers, and a fully connected layer for output. 

Next we detail the CNN design. Without loss of generality, we treat $W_{t-1}$ and $\hat{W}_t$ as a row vector, denoted as $w_{t-1}$ and $\hat{w}_t$, respectively, even though their original form can be matrix or tensor. This facilitates us performing dimension-wise operations on combining two models. \vspace{+5pt}

% design the transfer that can emphasize the relation between $w_{t-1}$ and $\hat{w_t}$ in the same dimension, but also capture the relation in the different dimension. 
%Inspired by human face recognition and other image classification tasks which capture local feature such as edge information in picture by CNN, We designed a CNN-based transfer. As shown in the green bloch of \ref{fig:overall-framework}, it has one concatenate layer, cnn layers, one full-connected layer, and an output layer. \\

\noindent\textbf{Stack layer}. 
%This layer, shown in the bottom of the green block of Figure~\ref{fig:overall-framework}, has two operations. (1) compute the element-wise product of $w_t$ and $\hat{w_t}$ \ie 
This layer stacks $w_{t-1}$, $\hat{w}_t$, and their element-wise product interaction vector as a 2D matrix, which serves as an ``image'' to be processed by the later convolution layers. Specifically, we formulate it as:
\begin{equation}
H^0 =
\begin{bmatrix}
w_{t-1} \\
\hat{w}_t \\
w_{dot} \\
%\end{bmatrix}, \text{where} \ w_{dot} = \frac{w_{t-1} \odot \hat{w_t}}{\|w_{t-1}\|^{0.5}\|\hat{w_t}\|^{0.5} + \epsilon}.
\end{bmatrix}, \text{where} \ w_{dot} = \frac{w_{t-1} \odot \hat{w_t}}{\|w_{t-1}\| + \epsilon}.
\end{equation}
The $w_{t-1} \odot \hat{w_t}$ can capture that when $w_{t-1}$ evolves to $\hat{w}_t$, which dimension values are enlarged or diminished. The denominator of $w_{dot}$ is used for normalization, and $\epsilon = 10^{-15}$ is a small number to prevent the denominator being zero. 
The size of $H^0$ is $3\times d$, where $d$ denotes the size of $w_{t-1}$ and $\hat{w}_t$. \vspace{+5pt}

\noindent\textbf{Convolution layers.} $H^0$ is fed into two cascaded convolution layers that further model dimension-wise relations. We describe the first convolution layer since the second one is formulated similarly. 
Let the first convolution layer have $n_1$ vertical filters, where each filter is denoted as $F_j \in \mathbb{R}^{3\times 1}$ (where $j = 1, ..., n_1$ denotes the filter index). $F_j$ slides from the first column to the last column of $H^0$ to perform operations on each column vector: 
\begin{equation}
	H_{j, m}^1 = \text{GELU} ( <F_{j} \ , \   H_{ : , m}^0 >), 
\end{equation}
where $H_{:,m}^0$ is the $m$-th column vector of $H^{0}$, $<  ,  >$ denotes vector inner product, and $H^{1}_{j,m} \in \mathbb{R}$ is the convolution result of $F_{j}$ on $H^{0}_{:,m}$. GELU is the Gaussian Error Linear Units activation function~\cite{GELU}, which can be seen as a smoothed variant of ReLU with gradients for negative values. 

Note that the vertical filter $F_j$ can learn various relations between $\hat{w}_t$ and $w_{t-1}$  at the same dimension. For example, if the filter is $[-1,1,0]$, it can express the difference between $\hat{w}_t$ and $w_{t-1}$; if the filter is $[1,1,1]$, it can obtain prominent features that have high positive value on both $\hat{w}_t$ and $w_{t-1}$. Another reason we use such 1D filter rather than the standard 2D filters is that the dimension order in $w_{t-1}$ or $\hat{w}$ is not meaningful for many recommender models. For example, if we permutate the embedding order for factorization models, the model prediction will not be changed. 

The output of the first convolution layer $H^1$ is a matrix of size $n_1 \times d$, which is then fed into the second convolution layer of $n_2$ filters, where each filter is of size $n_1 \times 1$. 
As a result, we obtain the output of this component $H^2$, which is a matrix of size $n_2 \times d$. \vspace{+10pt}

\noindent\textbf{Full-connected and output layers.} $H^2$ is fed into a fully-connected (FC) layer to capture the relations among different dimensions. We first flatten $H^2$ as a vector which has the size $d n_2$, and then feed into a fully connected layer:
\begin{equation}
	z = \text{GELU} ( W_f^T \text{flatten} (H^2) + b_1 ),
\end{equation}
where $W_f\in \mathbb{R}^{ (d n_2) \times d_f } $ and $b_1 \in \mathbb{R}^{d_f}$ are the weight matrix and bias vector of the FC layer, respectively, and $d_f$ denotes the layer size. The vector $z \in  \mathbb{R}^{d_f}$ is then transformed by a linear layer to output the new parameter vector $w_t$:
\begin{equation}
	w_t = W_o^T z + b_2, 
\end{equation}
where $W_o\in \mathbb{R}^{d_f \times d } $ and $b_2 \in \mathbb{R}^{d}$ are the weight matrix and bias vector of the linear layer, respectively. Lastly, the parameter vector $w_t$ is reshaped to the $W_t$ --- i.e., the new model parameters after retraining. 

To summarize, all trainable parameters of the transfer component are $\Theta = \{ F^{(1)}, F^{(2)}, W_f, b_1, W_o, b_2 \}$, where $F^{(1)} \in \mathbb{R}^{n_1 \times 3} $ and $F^{(2)} \in \mathbb{R}^{n_2 \times n_1}$ denote the filters of the first and second convolution layer, respectively. It is worth mentioning that 
we can categorize the parameters of a recommender model into different groups, and apply a separate transfer network for each group. For example, the matrix factorization model has two groups of parameters --- user embedding and item embedding. Then we use two transfer networks, one for user embedding and another for item embedding (see details in Section \ref{ss:method_mf}).

%%%%%%%%%%%%%%%%%%%%%%%%%%%%%%%%%%%%%%%%%%%%%%%%%%%%%%%%%%%%%%%%%%%%%%%%%%%%%%%%%%%%%%%%%%%%%%%%%%%%%%%%%%%%%%%%%%%%%%%%%%%%%%%%%%%%%%%%%
%      Training
%
%
\subsection{Sequential Training}

%\begin{figure} 
%	\centering
%	\includegraphics[width=0.5\textwidth]{picture/trainingprocess.pdf}
%	\caption{\textbf{One time period SML training process.}  In the figure any module with black color is fixed, white module is activate and will be learned. In the step 1, we fixed transfer module, to train new recommendation $\hat{W}$.In the step 2, we train transfer with fixed $\hat{W_t}$.} 
%	\Description{overall }
%	\label{fig:sml-training}
%\end{figure}

We now consider how to train model parameters, including the transfer input $\hat{W}_t$ for each task $\tau_t$, and the tranfer parameter $\Theta$ that is shared for all tasks. 
Functionally speaking, $\hat{W}_t$ is expected to extract recommendation knowledge from the current data $D_t$, whereas $\Theta$ combines the previous model $W_{t-1}$ and $\hat{W}_t$, which is expected to make the transfer output $W_t$ perform well on future data $D_{t+1}$. 
Since the data comes in sequentially, we perform training in the same sequential way, i.e., solving the task $\tau_{t-1}$ before moving to the next task $\tau_{t}$. 
Algorithm \ref{alg:SML} shows the sequential training process. We next describe how to train for a task $\tau_t$ (i.e., line 3 to 11), which has two main steps: \vspace{+5pt}

\noindent \textbf{Step 1. Learning the transfer input $\hat{W_t}$.} A straightforward solution is to directly learn it based on the recommendation loss on $D_t$. However, the resultant $\hat{W_t}$ may not be suitable as the input to the transfer, which assumes $W_{t-1}$, $\hat{W_t}$, and $W_t$ are in the same space (i.e., the parameter dimensions are aligned for the same semantics and the values are in the same scale range). To address this problem, we propose to optimize the transfer output on $D_t$, back-propogating gradients to the transfer input $\hat{W}_t$. Specifically, we formulate the loss as:
\begin{equation}\label{eq:loss-r}
L_r(\hat{W}_t | D_t) = L_0 ( f_\Theta(W_{t-1}, \hat{W}_t) | D_t) + \lambda_1 ||\hat{W}_t ||^2,
\end{equation}
where $L_0 ( x | D_t)$ denotes the recommendation loss (e.g., the log loss~\cite{NCF} or pairwise loss~\cite{BPR}) on data $D_t$ with $x$ as the recommender model parameters (note $x = f_\Theta(W_{t-1}, \hat{W}_t)$ here). $\lambda_1$ is a hyper-parameter to control $L_2$ regularization to prevent overfitting. 
When optimizing the loss, $\Theta$ is treated as constant and is not updated, so only the gradient of $\hat{W}_t$ needs be evaluated:
\begin{equation}\label{eq:gradient-r}
	\frac{\partial L_r(\hat{W}_t | D_t)}{\partial \hat{W}_t} = \frac{\partial L_0( x | D_t)}{\partial x} \cdot \frac{\partial f_\Theta(W_{t-1}, \hat{W}_t)}{\partial \hat{W}_t} + 2\lambda_1 \hat{W}_t, 
\end{equation}
%todo
After getting this gradient, we can apply gradient descent optimizer to update $\hat{W_t}$ like SGD and Adam~\cite{adam}. Through this way, we can achieve the two effects simultaneously 1) distilling recommendation knowledge from $D_t$, and 2) making $\hat{W}_t$ suitable as the input to the transfer network.

\vspace{+5pt}
\noindent\textbf{Step 2. Learning the transfer parameter $\Theta$.} Since $\Theta$ is shared across all tasks, it can capture some task-invariant patterns, e.g., which parameter dimensions are more relective of user short-term  interests and should be emphasized when combining $W_{t-1}$ and $\hat{W}_t$. The general aim is to obtain such patterns that are tailored for the next-period recommendations. As such, we consider optimizing $\Theta$ on the next-period data $D_{t+1}$. Specifically, we formulate the objective function as:
\begin{equation}\label{eq:loss-s}
L_s(\Theta | D_{t+1}) = L_0 ( f_\Theta(W_{t-1}, \hat{W}_t) | D_{t+1}) + \lambda_2 ||\Theta ||^2,
\end{equation}
where $\lambda_2$ is regularization hyper-parameter. Note that the $\hat{W}_t$ gained in Step 1 is a function of $\Theta$. Thus, when computing the gradients of $\Theta$, it will cause high-order gradients that are expensive to obtain. As such, we follow the first-order MAML algorithm~\cite{finn2017MAML}, ignoring such high-order gradients which have minor impacts on gradients but are expensive to obtain. 
By treating $\hat{W}_t$ as constant in this step, we evaluate the gradient of $\Theta$ as:
\begin{equation}\label{eq:gradient-s}
	\frac{\partial L_s(\Theta | D_{t+1})}{\partial \Theta} = \frac{\partial L_0( x | D_{t+1})}{\partial x} \cdot \frac{\partial f_\Theta(W_{t-1}, \hat{W}_t)}{\partial \Theta} + 2\lambda_2 \Theta, 
\end{equation}
where $x = f_\Theta(W_{t-1}, \hat{W}_t)$ for brevity. \vspace{+5pt}

The above two steps are iterated until convergence or a maximum number of iterations is reached (line 4). 
As Algorithm \ref{alg:SML} shows, the update of $\Theta$ is not performed in the last training period $T$, since its next period data $D_{T+1}$ is not available in training. Note that we can run multiple passes of such sequential training on $\{D_t\}_{t=0}^T$, while we empirically find one pass is sufficient to obtain good performance, thus we train only one pass. 

It is worth mentioning that the parameter update procedure of the serving (evaluation) phase slightly differs. Algorithm \ref{alg:testing} shows how we perform model evaluation for testing (validation) with newly collected data $ D_{t+1}$. First, we use it to test the model $W_t$ that served the period $t+1$.  
Then, we need to update $\Theta$ and $\hat{W}_{t+1}$ with $D_{t+1}$, so as to obtain $W_{t+1}$ for serving next period (note that $W_{t+1} = f_{\Theta}(W_t, \hat{W}_{t+1})$). 
As shown in line 3-8, we first iterate the updating of $\Theta$ and $\hat{W}_t$, which is same as the training phase. When stopping condition meets, we used the refreshed $\Theta$ to update $\hat{W}_{t+1}$, which is finally fed into $f_\Theta(W_t, \hat{W}_{t+1})$ to achieve $W_{t+1}$. 

\begin{algorithm}[t]
	\caption{Sequential Training of SML}
	\LinesNumbered
	\label{alg:SML}
	\KwIn{Training data of $T$ periods $\{D_t \}_{t=0}^T$}
	\KwOut{Recommender $W_T$, transfer $\Theta$}
	Randomly initialize $W_{-1}$ and $\Theta$ \;
    \For{$t=0$ to $T$}{
    	$\hat{W}_t \leftarrow W_{t-1} $ \;
    	\While{Stop condition is not reached}{
    	// Step 1: Learning $\hat{W}_t$ \\
    	Update $\hat{W}_t$ by optimizing $L_r(\hat{W}_t | D_t)$\;
    	// Step 2: Learning $\Theta$ \\
    	\textbf{if} $t == T$ \textbf{then} break \;  
    	Update $\Theta$ by optimizing $L_s(\Theta|D_{t+1})$\; 
		}
	    $W_t \leftarrow f_\Theta(W_{t-1},\hat{W}_{t})$ \;
    }
	return $W_{T}, \Theta$
\end{algorithm}

% \begin{algorithm}[t]
% 	\caption{Model Evaluation of SML}
% 	\LinesNumbered
% 	\label{alg:testing}
% 	\KwIn{Testing data of $S$ periods $\{D_{t} \}_{t=T+1}^{T+S}$ }
% 	%\KwOut{Recommender $W_T$, transfer $\Theta$}
% 	Randomly initialize $W_{-1}$ and $\Theta$ \;
% 	\For{$t=T+1$ to $T+s$}{
% 		Use $D_{t}$ to evaluate the model $W_{t-1}$ \;
% 		Update $\Theta$ with $D_t$, i.e., optimizing $L_s(\Theta | D_t)$  \;
		
% 		%Update $\hat{W}_{t}$ with $D_t$, i.e., optimizing $L_r(\hat{W}_t | D_t)$ \; 
% 	}
% \end{algorithm}

\begin{algorithm}[t]
	\caption{Model evaluation and update}
	\LinesNumbered
	\label{alg:testing}
	\KwIn{Newly collected data $D_{t+1}$, recommender $W_t$ to test }
	\KwOut{Updated recommender $W_{t+1}$}
	Use $D_{t+1}$ to test the model $W_{t}$\;
	// Model update for next period\; 
	\While{Stop condition is not reached}{
	Update $\Theta$ by optimizing $L_s(\Theta|D_{t+1})$ \;
	Update $\hat{W}_t$ by optimizing $L_r(\hat{W}_t | D_t)$ \;
	}
	Run line 4 and $W_t \leftarrow f_\Theta(W_{t-1},\hat{W}_{t})$ \;
	Update $\hat{W}_{t+1}$ by optimizing $L_r(\hat{W}_{t+1} | D_{t+1})$ \;
	$W_{t+1} = f_{\Theta}(W_t, \hat{W}_{t+1})$ \;
	return $W_{t+1}$
\end{algorithm}
%\vspace{-0.5cm}

% \begin{algorithm}
% 	\caption{one time period SML train process}
% 	\LinesNumbered
% 	\label{alg:one-stage-SML}
% 	\KwIn{ $W_{t-1},\Theta_{t-1},D_{t},D_{t+1}$}
% 	\KwOut{$W_t,\Theta_t$}
% 	init $\hat{W_t} = W_{t-1}$, $\Theta = \Theta_{t-1}$\;
% 	\While{not stop}{
% 		\For{batch in $D_t$}{
% 			compute $\widetilde{W}$ according to \eqref{equation:widetilde-W-1}\;
% 			compute loss $L_R(\hat{W_t};bacth)$ according to \eqref{equation:w-hat-loss} with the data bacth.\;
% 			compute gradient of $\hat{W_t}$ \wrt $L_R$ with fixed $\Theta$ \ie $g(\hat{W_t};batch)$, according to \eqref{equation:W-hat-gradient}\;
% 			Update $\hat{W_t}$ with  $g(\hat{W_t};batch)$ by SGD.
% 		}
% 		\If {first outer loop and need test}{
% 			$W_{t} = f(\hat{W_t},W_{t-1};\Theta)$ \;
% 			test $D_{t+1}$ on $W_{t}$
% 		}
		
% 		\For{batch in $D_{t+1}$}{
% 			compute $\widetilde{W}$ according to \eqref{equation:widetilde-W-2}\;
% 			compute loss $L_R(\hat{W_t};bacth)$ according to \ref{equation:Theta-loss} with the data bacth.\;
% 			compute gradient of $\Theta$ \wrt $L_R$ with fixed $\Theta$ \ie $g(\Theta;batch)$, according to \eqref{equation:theta-gradient}\;
% 			update $\Theta$ with  $g(\Theta;batch)$ by SGD.
% 		}
% 		$\hat{W_t} = f(\hat{W},W_{t-1},\Theta)$ (optional)
% 	}
% 	$\Theta_{t} = \Theta; W_{t} = f(\hat{W},W_{t-1},\Theta)$\;
% 	return $\Theta,W_{t}$
% \end{algorithm}

\subsection{Instantiation on Matrix Factorization} \label{ss:method_mf}

To demonstrate how our proposed SML framework works, we provide an implementation based on matrix factorization (MF), a representative embedding model for recommendation. Given a user-item pair $(u,i)$, MF predicts the interaction score as:  
\begin{equation}
	\hat{y}_{ui} = p_u^T q_i,
\end{equation}
where $p_u \in \mathbb{R}^{dim}$ and $q_i \in \mathbb{R}^{dim}$ denote the embedding of user $u$ and item $i$, respectively, and $dim$ denotes the embedding size. 
As we can see, MF has two groups of parameters: user embedding and item embedding. Thus we build two separate transfer networks, one for user embedding and another for item embedding.  
%We share the user transfer network for all user embeddings, i.e., the input $w_{t-1}$ is an embedding of a user that has the size $dim$; same for the item transfer network. 
Instead of feeding the embeddings of all users into the user transfer network, we operate the transfer network on the basis of each user embedding; same for the item side. 
The rationality is that the semantics of embedding dimensions across all users are the same, thus we can share the transfer network for all users. 
This largely reduces the number of transfer parameters and makes the transfer more generalizable. 

For the recommendation loss $L_0$, we adopt the pointwise log loss, which is a common choice for recommender training~\cite{NCF,xDeepFM}. For each interaction $(u,i) \in D_t$, we randomly sample 1 unobserved interactions of $u$ to form the negative data set $D_t^-$. Then the log loss is formulated as:
\begin{equation}
	L_0(P, Q | D_t) = -\sum_{(u,i) \in D_t } \log (\sigma(\hat{y}_{ui})) - \sum_{(u,j) \in D_t^- } \log (1 - \sigma(\hat{y}_{uj})),
\end{equation}
where $P$ and $Q$ denote the embeddings of all users and items, $\sigma(\cdot)$ denotes the sigmoid function. The same log loss is used in both optimization of the recommender  $L_r$ and the transfer $L_s$. 
\section{EXPERIMENTS}\label{sec:experiment}
%\quad In this section, we present our experimental setup on two dataset with different properties.
We conduct experiments to answer the following questions:
\begin{itemize}
    \item[\textbf{RQ1:}] How is the performance of SML compared with existing retraining strategies and recommender models?
    \item[\textbf{RQ2:}] How do the components of SML affect its effectiveness?
    \item[\textbf{RQ3:}] How does the CNN architecture affect the transfer network?
    \item[\textbf{RQ4:}] Where are the improvements of SML come from?
\end{itemize}

We first present experimental settings, followed by results and analyses to answer each research question.

\begin{table*}[]
\caption{Average recommendation performance over online testing periods on Adressa and Yelp. ``RI'' indicates the relative improvement of SML over the corresponding baseline.}
\vspace{-0.4cm}
\label{tab:main-result}
\resizebox{0.90\textwidth}{!}{%
\begin{tabular}{clccclcccl}
\toprule%\hline
\multicolumn{1}{l}{Datasets} & Methods & recall@5 & recall@10 & recall@20 & RI & NDCG@5 & NDCG@10 & NDCG@20 & RI \\ \hline
\multicolumn{1}{c|}{\multirow{6}{*}{Adressa}} & Full-retrain & 0.0495 & 0.0915 & 0.1631 & 319.7\% & 0.0303 & 0.0437 & 0.0616 & 393.1\% \\%0.0469 & 0.0883 & 0.1601 & 336.9\% & 0.0285 & 0.0417 & 0.0597 & 417.7\% \\
\multicolumn{1}{c|}{} & Fine-tune & 0.1085 & 0.2235 & 0.3776 & 82.8\% & 0.0594 & 0.0962& 0.1351 & 135.5\% \\
\multicolumn{1}{c|}{} & SPMF & 0.1047 & 0.2183 & 0.3647 & 87.3\% & 0.0572 & 0.0935 & 0.1306 & 143.6\% \\
\multicolumn{1}{c|}{} & GRU4Rec & 0.0213 & 0.0430 & 0.0860 & 809.0\% & 0.0125 & 0.0194 & 0.0302 & 1018.4\% \\
\multicolumn{1}{c|}{} & Caser & 0.2658 & 0.3516 & 0.4259 & 6.5\% & 0.1817 & 0.2096 & 0.2285 & 2.1\% \\ \cline{2-10} 
%\multicolumn{1}{c|}{} & weSum &0.0936 & 0.1932 & 0.3306 & 111.1\% & 0.0517 & 0.0836 & 0.1183 & 170.3\% \\ \cline{2-10} 
\multicolumn{1}{c|}{} & \textbf{SML} & \textbf{0.2815} & \textbf{0.3794} & \textbf{0.4498} & \multicolumn{1}{c}{-} & \textbf{0.1838} & \textbf{0.2156} & \textbf{0.2336} & \multicolumn{1}{c}{-} \\ \hline
\multicolumn{1}{c|}{\multirow{6}{*}{Yelp}} & Full-retrain & 0.1849 & 0.2876 & 0.4139 & 18.0\% & 0.1178 & 0.1514 & 0.1829 & 22.7\% \\
\multicolumn{1}{c|}{} & Fine-tune & 0.1507 & 0.2386 & 0.3534 & 41.7\% & 0.0963 & 0.1246 & 0.1535 & 48.5\% \\
\multicolumn{1}{c|}{} & SPMF & 0.1664 & 0.2591 & 0.3749 & 30.7\% & 0.1072 & 0.1370 & 0.1662 & 35.1\% \\
\multicolumn{1}{c|}{} & GRU4Rec & 0.1706 & 0.2764 & 0.4158 & 22.8\% & 0.1080 & 0.1420 & 0.1771 & 30.5\% \\
\multicolumn{1}{c|}{} & Caser & 0.2195 & 0.3320 & 0.4565 & 2.8\% & 0.1440 & 0.1802 & 0.2117 & 3.12\% \\ \cline{2-10} 
% \multicolumn{1}{c|}{} &weSum&0.1571 & 0.2401 & 0.3461 & 40.4\% & 0.1039 & 0.1306 & 0.1573 & 41.3\% \\\cline{2-10} 
\multicolumn{1}{c|}{} & \textbf{SML} & \textbf{0.2251} & \textbf{0.3380} & \textbf{0.4748} & \multicolumn{1}{c}{-} & \textbf{0.1485} & \textbf{0.1849} & \textbf{0.2194} & \multicolumn{1}{c}{-} \\ \bottomrule%\hline
\end{tabular}%
}
\vspace{-0.1cm}
\end{table*}

\subsection{Experimental Settings}
\subsubsection{Datasets} 
We experiment with two real-world datasets from Adressa and Yelp. 
%which contains the sequential data for item recommendation in different business scenarios.
%The statistics of the datasets  is shown in Table1. Table~\ref{tab:datasets}.

\textbf{Yelp:} The dataset is adopted in Yelp Challenge 2019\footnote{https://www.yelp.com/dataset/}, which contains the interaction records between users and businesses like restaurants and bars, spanning a period of more than 10 years. For ease of evaluation, we remove the inactive users with less than 10 interactions and unpopular items with less than 20 interactions. The experimented data contains 3,014,421 interactions from 59,082 users and 122,816 items.

\textbf{Adressa~\cite{adressa}:} 
%The dataset is a news dataset that includes news articles in connection with anonymized users. It recorded 3,664,225 clicks by 478,612 people in three weeks. We treat every click on the article as a positive sample.
The dataset is from Adressa\footnote{http://reclab.idi.ntnu.no/dataset/}, which records user clicks on news articles in three weeks. %To ensure the quality of dataset, we filter out incomplete records with active-time of zero, 
We remove invalid interactions that have a news reading time of zero. The experimented data has 3,664,225 interactions between 478,612 users and 20,875 items.

%For each dataset, we split it into two parts, one is for offline training denote as $D_{offline}$ and the other is for online denote as $D_{online}$. 
We purposefully choose the two datasets because of their different properties --- the Adressa dataset emphasizes more on user short-term interest, since the news domain is more time-sensitive; in contrast, the Yelp dataset emphasizes more on user long-term interest, since it lasts longer and a user's choice on businesses is less time-sensitive. 

To test the periodical model retraining, we organize each dataset into periods (\ie $\{D_{0},\dots, D_{T}\}$) according to the timestamp of interaction. For Adressa, we split each day into three periods based on the morning (0:00-10:00), afternoon (10:00-17:00), and evening (17:00-24:00), obtaining 63 periods in total. 
As for the Yelp dataset which lasts longer, we split it into 40 periods with an equal number of interactions, where each period roughly corresponds to a quarter. 
%We respectively hold out the last 15 periods and 10 periods of Adressa and Yelp for online testing. 
We further split the periods of each dataset into training/validation/testing sets: for Adressa the ratio is 48/5/10, and for Yelp the ratio is 30/3/7. For each testing period, data collected in all the previous periods can be used for model retraining.

\subsubsection{Baselines}
% We compare our methods with three variants of Matrix Factorization, a sample-based method, and two state-of-the-art sequential recommendation method. To make a more convincing comparison with the baseline, all methods use Binary Cross Entropy Loss as loss function. For all the baselines, we pre-train initial models based on $\{D_{0},\dots,D_{47}\}$ of Adressa, and $\{D_{0},\dots,D_{29}\}$ of Yelp. We start the first period training with t=48 on Adressa and t=30 on Yelp. 
We compare the proposed SML method with three \textit{retraining strategies} that are also applied to MF:

\textbf{- Full-retrain.} This method trains the MF model on all past data $\{D_0, ..., D_{t-1}\}$ and newly collected data $D_t$ at each period $t$. %\textbf{- Full Training MF}: Matrix Factorization is a classic model for personalized recommendation, which use the inner product of user embedding and item embedding as the metrics of interest.For this experiment, We use all historical information before the test period. That is, use $\{D_{1},D_{2},\dots,D_{t}\}$ as the training set while test on $D_{t+1}$.

%\textbf{- Fine Tuning MF.} Fine-tune \ie retraining model only with the current time data $D_{t}$, is an other common strategy. We applied it to MF. %Based on MF model, use only $D_t$ as training set when test on $D_{t+1}$. Furthermore, we train on the basis of the parameters in the previous period.
\textbf{- Fine-tune.} This method updates the MF model on the newly collected data $D_{t}$ only. 

\textbf{- SPMF~\cite{SIGIR18-YinHongzhi}.} This is a state-of-the-art streaming recommendation method that belongs to the category of sampled-based retraining. It maintains a reservoir of historical interaction samples and adds them into $D_{t}$ to retrain the MF model. 
We tune the reservoir size in \{7000,15000,30000,70000\}.  % It develops a reservoir to represent user historical interest by statistical and update the reservoir over time. We set the reservoir size from {7000,15000,30000,70000,150000}.\\

We also compare with two \textit{sequential recommendation} methods, which are designed for modeling sequential user-item interactions:

%\textbf{- GRU4Rec\cite{GRU4Rec}.} GRU4Rec is not a model for retraining, but a GRU-based sequential model for session-based recommendation. It take a session of a user into the GRU, and recommend the user the next items. Because there is no session information for Yelp and Adressa in our setting, we treat all the history interactions of a user as a session . For this reason, we need retrain it with $\{D_{0},\dots,D_{t}\}$ in each time period t. The hidden layer of GRU is select from $\{64,128,256\}$. %A RNNs based method that uses user interaction sequence for session-based recommendation. We treat all of one’s interaction in $\{D_{1},D_{2},\dots,D_{t}\}$  as a session. The hidden layer of GRU is select from $\{64,128,256\}$.
\textbf{- GRU4Rec~\cite{GRU4Rec}.} This is a representative sequential recommender based on recurrent neural network (RNN). It builds a RNN for each user's interaction sequence to capture her interest evolution. We employ full retraining strategy at each testing period, which performs better than fine-tuning for the method, and keep loss as the paper. The hidden layer size of GRU is tuned in the range of $\{64,128,256\}$.

\textbf{- Caser~\cite{Caser}.} This method uses CNN for sequential modeling. It takes $L$ most recently interacted items and forms their embeddings as a 2D matrix, feeding the matrix into a CNN with two types of convolution layers --- horizontal layer and  vertical layer. We tune $L$ in the range of $ \{2,...,5\}$,CNN kernel number in $ \{4,8,16\}$, and other hyper-parameters follow the optimal setting as reported in the paper. We employ fine-tuning strategy at each testing period, which performs better than full training for the method. 

For fair comparison, all methods are optimized with the same log loss (except GRU4Rec) and tuned on the validation set. 
For the four methods based on MF (\ie Full-retrain, Fine-tune, SPMF, and our SML), we tune three hyper-parameters: $L_2$ regularization coefficient  $\lambda$ in $\{1e\text{-}1,1e\text{-}2,...,1e\text{-}7,0\}$, learning rate in $\{0.1,0.01,0.001\}$, and training epochs in \{5,10,20,50,100\}, respectively.
%\{1e-1,1e-2,...,1e-6,0\}, lr(learning rate) among {0.1,0.01,0.001}, and update epochs among \{10,20,50,100,200\}.
For SML, we additionally tune the number of maximum iterations (line 4 of Algorithm 1) in $\{5,6,\dots,10\}$. 
%We initialize our MF model by pre-training bon $\{D_{0},\dots,D_{20}\}$ of Adressa and $\{D_{0},\dots,D_{9}\}$ of Yelp. For the first training period, t is equal to 10 on Yelp and 21 on Adressa.
%Based on the recommendation performance on the validation data evaluated by recall@10, we set the hyper-parameters of SML as follows:  $\lambda$=1e-4,$lr$=1e-3 for training $\theta$, $\lambda$=1e-6,$lr$=1e-2 for training $\hat{W_t}$. 
The CNN filter size is set as $[10, 5]$ and the fully connected layer size is $512$ for both datasets. 
%For our transfer model, we set CNN filters as [10,5], FC layers as [512,512,64] and the number of alternate updates as 7 for Adressa ,10 for Yelp.

%For baselines except for linear transfer, when offline, we train a initial model. At online stage, we use the first 1/3 periods to find stop condition and hyper-parameters which need be Fine-tune for retraining, and test the performance in the last 2/3 periods. For SML and linear transfer, offline training follows the process in Algorithm 1, first 1/3 periods of handout periods are treat as validation to find hyper-parameters and stop condition.

\subsubsection{Evaluation Protocols}
To simulate the real-world scenario that there are typically some historical data to train an initial model, we start model retraining from the 10-th and 20-th period of Yelp and Adressa, respectively, using the previous data to train an initial model. 
We perform evaluation at each testing period and report the average scores.
The evaluation is done on each interaction basis. 
As it is time consuming to rank all non-interacted items, we sample $999$ non-interacted items of a user as the recommendation candidates. 
For each testing interaction, the method outputs a ranking list on the 1 interacted item and $999$ non-interacted items. 
%Then, each method outputs the user's preference scores over all positive and negative items. 
We adopt two widely-used evaluation metrics: Recall@K and NDCG@K~\cite{NCF} and $K$ is set to 5,10, and 20. For parameters tuning on validation sets, we take Recall@20 as the main referential metric. 

\subsection{Performance Comparison (RQ1)}
% We now compare our transfer with baselines. 
\subsubsection{Overall Comparison.}
Table~\ref{tab:main-result} shows the top-$K$ recommendation performance of compared methods. From the table, we have the following observations:
\begin{itemize}[leftmargin=*]
    \item Full-retrain outperforms Fine-tune on Yelp, but it is significantly worse on Adressa. This shows the varying properties of the two datasets: Yelp users choose businesses based more on their inherent (long-term) interest, whereas Adressa users are more time-sensitive to choose recent news and driven by their short-term interest. 
    %Since Full-retrain considers user long-term interest and Fine-tune emphasizes short-term interest, this demonstrates the 
    %The performance of Full-retrain that highlights the long-term user preference is better than Fine-tune that emphasizes the short-term user preference on Yelp, but worse on Adressa.% It seems not impossible that training a model with more previous data will cause a worse result.
    %The result is reasonable since Adressa is a news dataset where the timeliness is very strong and the short-term interest is thus more important;   %We shows this by a simple experiments, we recommend user most-popular items from items exist w days ago till now, and the results are shown in Table~\ref{}. We find a small w is, a better result we have.  
    %Yelp is a dataset of local business where items may exist several years and long-term interest thus could play a more important role. 
    Full-retrain makes use of all data to do model retraining, which is effective in capturing user long-term interest; however, it suffers from emphasizing the importance of recent data. 
    This demonstrates the necessity of properly handling both long-term and short-term user preference in the periodical model retraining.
    % \item The proposed SML achieves the best performance on the two datasets. Besides the Carser, our methods has gotten relative improvement more than 20\% in yelp and more than 80\% in Adressa. Compared with Caser, we can get an average relative improvement up to 6\% on recall and 2\% on NDCG for Adressa, and can get about 3\% RI for yelp. It's amazing that retraining methods can get this performance only based on MF model. This shows the the power of our model.
    \item Our proposed SML achieves the best performance on both datasets, consistently outperforming Full-retrain and the most competitive baseline Caser. 
    %Statistical tests verify all improvements are significant. 
    This result signifies the effectiveness of SML, which is attributed to the dedicated design of the transfer network and the sequential training algorithm. The strong performance on both datasets shows that SML is capable of adapting long-term and short-term interests by optimizing the transfer network on the next-period data.
    \item In particular, SML outperforms Full-retrain by 18\% on Yelp. It validates our belief that historical data can be discarded during retraining, as long as the previous model can be properly utilized. This can largely save computation resources in model retraining, which has high value for practical use.  
    %It validates the rational of our reformulation of the retraining process in Equation~\ref{2}, \ie, historical data could be discarded in model retraining as long as the previous model is properly utilized, which is of high practical value for saving computation resources. 
    %\item Moreover, the proposed SML outperforms Fining-tuning on Adressa by 106.15\% on average. This improvement is attributed to the Transfer model which is able to adjust the model on the new interactions so as to get good recommendation in the near future. It also validates the effectiveness of viewing next period performance as one of the optimizing objectives of model retraining.
    
    %\item However, weSum that implements the Transfer model as a linear combination of the current model and previous model performs worse than Fine-tune on both datasets. This result shows that the relation between the knowledge in the previous model and the current model might not be a shallow linear one, suggesting the usage of a learnable and more complex transfer.
    
    \item The sample-based retraining method SPMF performs better than Fine-tune on Yelp, but not on Adressa. The reservoir in SPMF is designed heuristically to bias towards retaining old interactions, which shows strength in capturing long-term user interest. However, it sacrifices the ability of modeling short-term interest, making it fall short on recommendation scenarios where recent data are more important. Generally speaking, such sample-based retraining method pursues a trade-off between Fine-tune and Full-retrain, and its performance is bounded by either method (on Adressa SPMF is weaker than Fine-tune and on Yelp SPMF is weaker than Full-retrain). 

    \item Among the two sequential recommender models, Caser performs much better than GRU4Rec, which implies that CNN would be a better choice than RNN to model the interaction sequence. Moreover, Caser outperforms the three MF-based baselines, which indicates its effectiveness in sequence modeling. However, its advantages can be surpassed by our SML, which wisely retrains MF towards the next-period performance. As a future extension, we will implement SML on Caser to see whether combing their advantages can lead to further improvements. 
    %Among the sequential recommendation baselines, Caser gets a strong performance which is significantly better the GRU4Rec. Noticing that Caser also utilizes CNN layers to encode the interaction sequence, we postulate that the CNN-structure is a wise choice to model the evolution of user preference. %Moreover, Caser also outperforms the conventional retraining strategies, showing the importance of con
\end{itemize}

% \begin{figure}[H] 
% 	\centering
% 	\subfigure[Adressa]{
%     \begin{minipage}[t]{0.23\textwidth}
%     \includegraphics[width=4cm]{picture/main-periods-Adressa.pdf}
%     \end{minipage}%
%     }
%     \subfigure[Yelp]{
%     \begin{minipage}[t]{0.23\textwidth}
%     \includegraphics[width=4cm]{picture/main-periods-yelp.pdf}
%     %\caption{fig1}
%   \end{minipage}%
%   }
% 	\caption{\textbf{Ablation study.} Show the function of different parts of our methods.}
% 	\Description{ablation }
% 	\label{fig:ablation}
% \end{figure}

% \begin{figure}[t]
% 	\centering
% 	\includegraphics[width=0.48\textwidth]{picture/main-period_new.pdf}
% 	\vspace{-1.0cm}
% 	\caption{Recall@10 of each testing period.} \vspace{-10pt}
% 	%\Description{mian result }
% 	\label{fig:main-result-period}
% \end{figure}
\begin{figure}[t]
	\centering
	\includegraphics[width=0.48\textwidth]{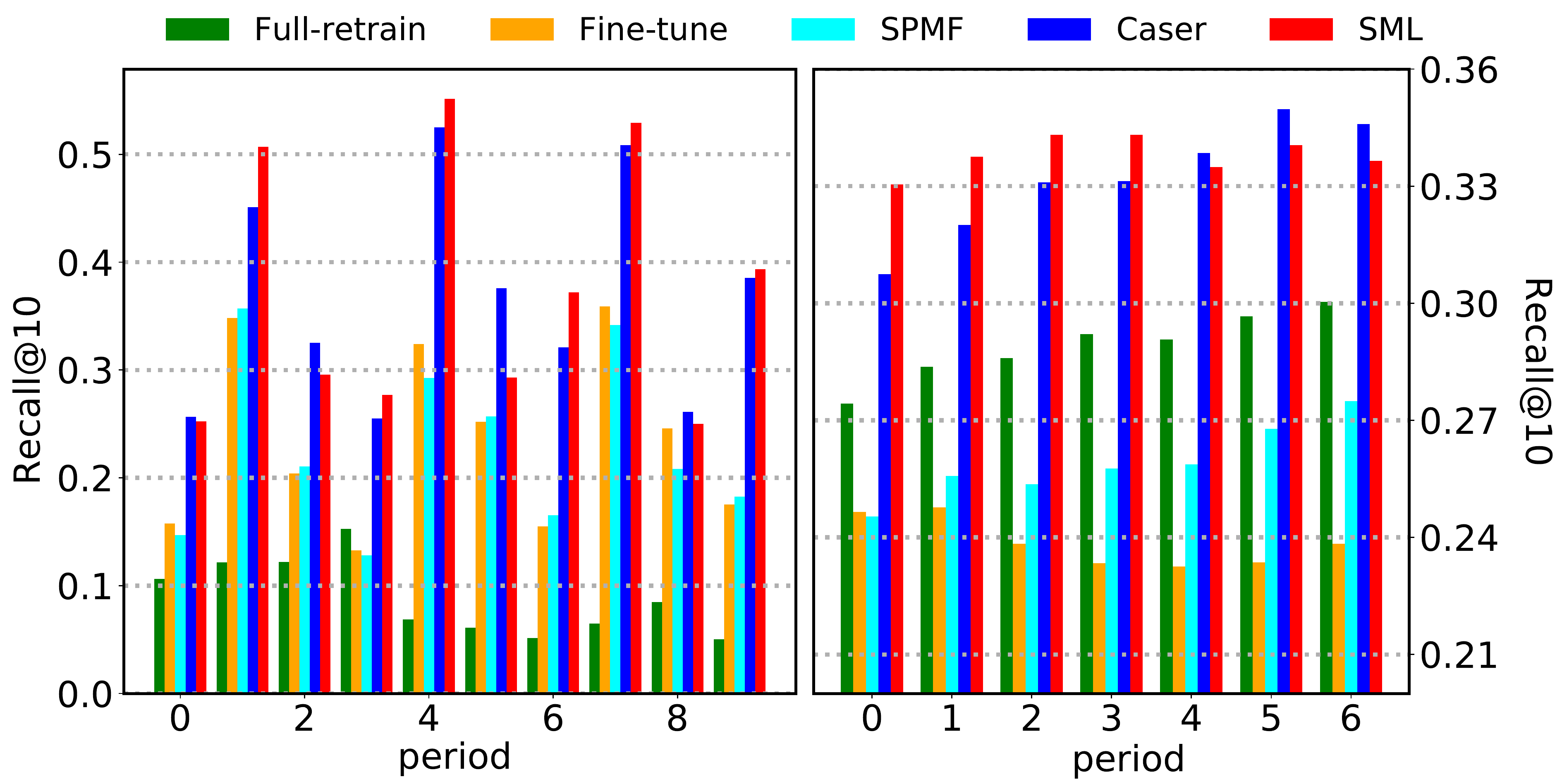} % 0521 main-period_new.pdf
	\vspace{-0.8cm}
	\caption{Recall@10 of each testing period.}
	\vspace{-10pt}
	%\Description{mian result }
	\label{fig:main-result-period}
\end{figure}
\subsubsection{Period-wise Performance.}
Figure~\ref{fig:main-result-period} shows the detailed recommendation performance at each online testing period evaluated by recall@10. To save space, we omit the results of other metrics which show the same trend\footnote{Since GRU4Rec achieves the lowest scores, its results are not shown in the figure for better visualization.}. From the figure, we can see that
%1) For Adressa, our SML achieves optimal results in almost all stages. But for yelp, SML has a worse results compared with Caser. 
SML achieves the best performance in most cases, which further validates its strong generalization ability. Moreover, the fluctuations on Adressa are larger than Yelp, which further validate the strong timeliness of the news domain (\ie user interest changes quickly) and the importance of performing fast model retraining so that the recommender is adapted to the changes of short-term interest. 
%because the news domain is more time sensitive. 
%2) Compared with Yelp, all the performance of Adressa is unstable at different stages. And it fluctate with a period of 3, this implies tah the unstable may be caused by the characteristics of the data. % xxxx. 2) xxxx. 3) xxxx.
%2) Compared with Yelp, regarding SML, the performance across different periods of Adressa fluctuate in a wide range with a period of 3. The result indicates the ability of SML to tolerate the fluctuation of historical model and current model. It validates the robustness of SML which is attributed to the sequential training algorithm which updates the parameter of transfer model on the fly.

\begin{table}[t]
\caption{Retraining time (seconds) at each testing period on Yelp. SML-S is the variant that disables transfer update.}\vspace{-13pt}
\label{tab:time_cost}
\begin{tabular}{@{}llllllll@{}}
\toprule
period       & 0   & 1    & 2    & 3    & 4    & 5    & 6    \\ \midrule
%Full-retrain & 978 & 1,011 & 1,153 & 1,191 & 1,262 & 1,282 & 1,328 \\
Full-retrain & 1,458 & 1,492 & 1,546 & 1,599 & 1,634 & 1,701 & 1,749 \\
% SML          & 65  & 63   & 62   & 61   & 61   & 61   & 64   \\\hline
SML          & 90  & 91   & 89   & 89   & 89   & 89   & 90   \\\hline
% Fine-tune    & 18  & 20   & 18   & 22   & 18   & 18   & 20   \\
Fine-tune    & 34  & 34   & 35   & 34   & 34   & 35   & 34   \\
SML-S        & 8   & 8    & 8    & 8    & 8    & 8    & 8    \\ \bottomrule
\end{tabular}
    %%%%%
    \vspace{-1pt}
	\justify
	\scriptsize{Full-retrain is trained with 20 epochs with recommendation performance slightly worse than that reported in Table 1 which is trained with 100 epochs.}
	%%%%%
	\vspace{-0.2cm}
\end{table}
% \begin{figure}[] 
% 	\includegraphics[width=0.45\textwidth]{picture/time-cost.pdf}
% 	\vspace{-0.5cm}
% 	\caption{Retraining time (seconds) at each period of Yelp.} 
% 	\Description{time cost}
% 	\label{fig:time_cost}
% \end{figure}
\subsubsection{Speed-up.} Recall that one motivation of the work is to accelerate model retraining by avoiding using previous data. 
We compare the retraining time of SML with Full-retrain and Fine-tune at different testing periods. 
The testing platform is a 1080Ti GPU with 2 CPUs and 16GB memory.
%, and the SML-s (\ie not update transfer when testing periods, more details in section~\ref{sec:experiment:ablation}). 
Table \ref{tab:time_cost} shows the time cost on Yelp by period. As can be seen: 1) the time cost of Full-retrain increases linearly as the testing process goes on, which is caused by the increase of training data. 2) SML is about 18 times faster than Full-retrain, and the retraining time is stable across different periods. 3) By disabling the update of the transfer network in the testing process, SML-S is even faster than Fine-tune, and also outperforms Fine-tune in recommendation accuracy (see Figure \ref{fig:ablation}). This shows the potential of SML in supporting fast model retraining, which is highly valuable in practice. 
% It validates the practical value of the proposed SML.

%\begin{figure} 
%	\includegraphics[width=0.3\textwidth]{picture/time.pdf}
%	\caption{Recall@10 of Yelp on last 10 periods.} 
%	\Description{over time }
%	\label{fig:over time}
%\end{figure}

\subsection{Ablation Studies (RQ2)} \label{sec:experiment:ablation}
% In this section, we will study the effect of different components in our method on performance. There are two main improvements. (1)We build an expressive component that  could transfer the historical knowledge from previous parameters to new coming data. (2)Our SML mechanism optimizes the transfer component towards future performance. To verify that our improvement worked, we design the following five variants and compare our method with them, and the results are shown in Fig4.     Figure~\ref{fig:ablation}. 
The strengths of SML come from two novel components: 1) the transfer network that combines the ``knowledge'' contained in the old model and new model; and 2) the sequential training process that optimizes the transfer network towards next-period performance. 
To justify the designs in SML, we investigate the influence of each important design. We study the performance of the following five variants: 

\noindent \textbf{- SML-CNN}, it removes the CNN layers from the transfer network.

\noindent \textbf{- SML-FC}, which removes the FC layer from the transfer network.

\noindent \textbf{- SML-N}, which disables the optimization of the transfer network towards the next-period performance. It trains all parameters on $D_t$, i.e., replacing $L_s(\Theta|D_{t+1})$ with $L_s(\Theta|D_{t})$ in Line 9 of Algorithm~\ref{alg:SML}.

\noindent \textbf{- SML-S}, which disables the update of the transfer network during testing, i.e., removing Line 4 of Algorithm~\ref{alg:testing}. The transfer network is fixed when updating the recommender. 

\noindent \textbf{- SML-FP}, which learns the transfer input $\hat{W_{t}}$ directly based on the recommendation loss on $D_t$, rather than forward propagation through the transfer network.

%Through the experimental results, we have the following observations.

\begin{figure} 
	\centering
% 	\subfigure[Adressa]{
%     \begin{minipage}[t]{0.24\textwidth}
%     \includegraphics[width=3.5cm]{picture/ablation_news_recall.pdf}
%     %\caption{fig1}
%     \end{minipage}%
%     }
%     \subfigure[Yelp]{
%     \begin{minipage}[t]{0.24\textwidth}
%     \includegraphics[width=3.5cm]{picture/ablation_yelp_recall.pdf}
%     %\caption{fig1}
%     \end{minipage}%
%     }
    \mbox{
		%\hspace{-0.1in}
		\subfigure[Adressa]{
		    \includegraphics[width=0.24\textwidth]{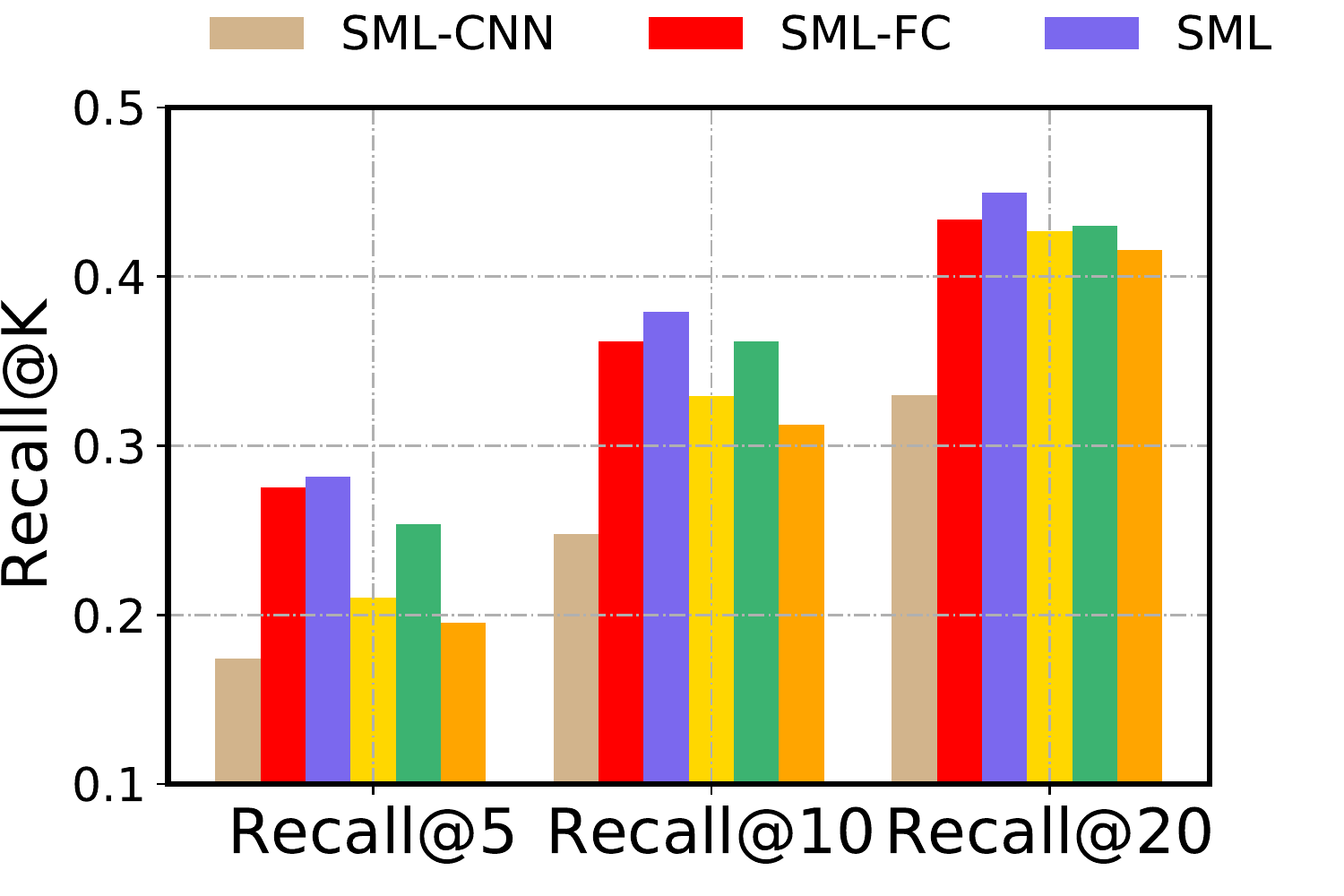}
        }
        \hspace{-0.2in}
		\subfigure[Yelp]{
		    \includegraphics[width=0.24\textwidth]{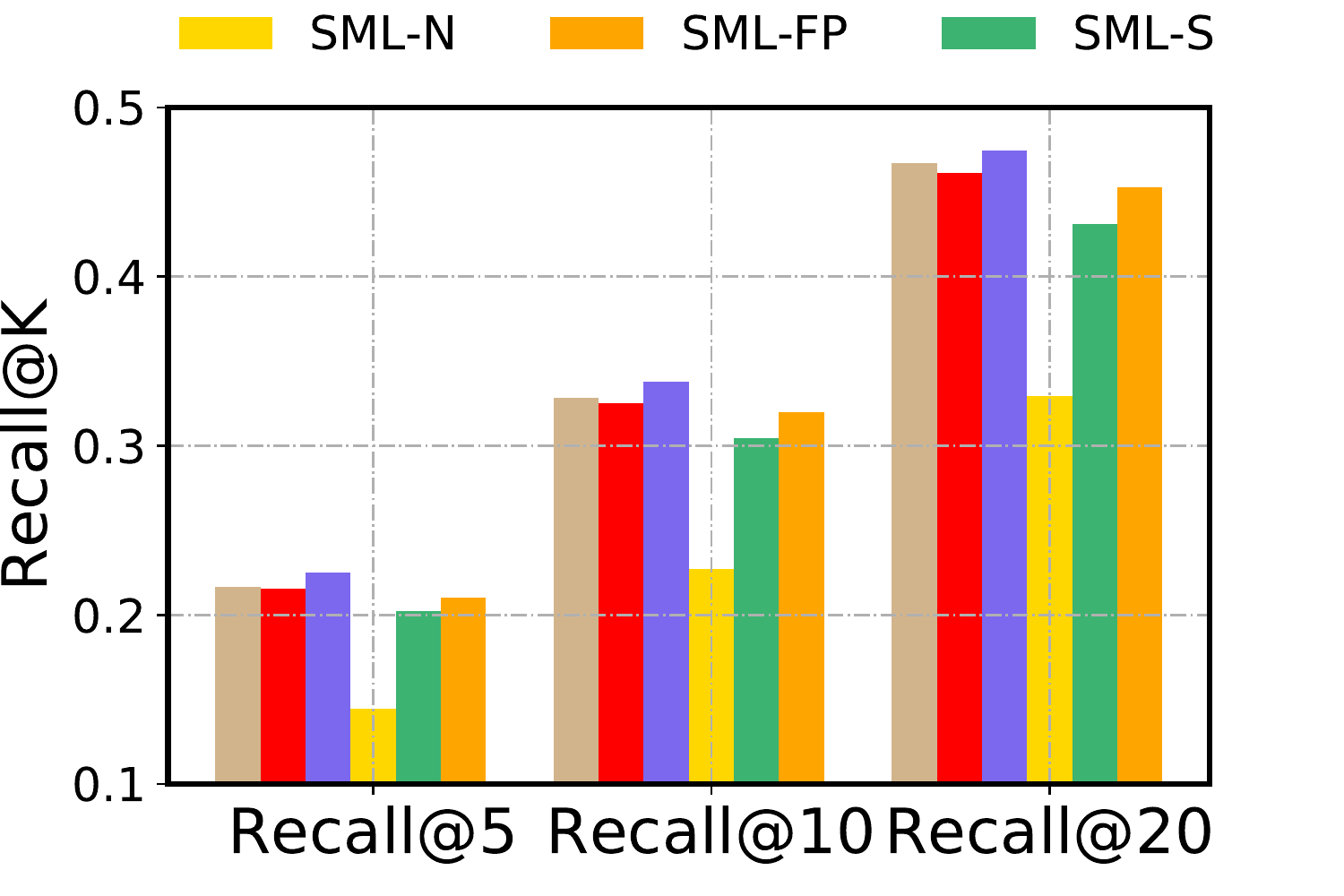}
        }
	}
	\vspace{-0.6cm}
	\caption{Recommendation performance of SML and its five variants with different designs being disabled.}
	\Description{ablation }
	\vspace{-16pt}
	\label{fig:ablation}
\end{figure}

%(1)CNN-based model performs much better than MLP-based model on Adressa. But, in the result of Yelp dataset, MLP-based model achieves similar performance with CNN-based model. That's to say, CNN structure can capture short-term interest evolution more easily than MLP. For both dataset, CNN-fc achieves excellent performance which shows that CNN layers work indeed.
%(2)According to the result of CNN-n,if only learn to meet the current data distribution, the model will achieve a very poor performance. Using future data makes it possible to learn how to transfer parameters to fit the data for the next period. Compared with result on Yelp, CNN-n have less decrease on performance of Adressa test. This is because Adressa's span is much shorter than Yelp. The longer the time span of the data, the greater the effect of our SML method. 
%(3)In CNN-load method, $\hat{w_{t}}$ is learned alone. Obviously, the performance is significantly reduced at this time. As mentioned earlier, the scale range of $\hat{w_{t}}$ may not suitable for current transfer without following the lead of transfer model.
%(4) We show Recall@10 of Yelp on last 10 periods of Retrain-MF, CNN-s, and CNN in Fig5.We can observe that if stop training Transfer model after test, the performance will decrease over time. Therefore, it's necessary to update Transfer model after test. Nevertheless, CNN-s can still outperform Retrain-MF for a long time.
Figure~\ref{fig:ablation} shows the recommendation performance of SML and the five variants on Adressa and Yelp. We omit the results of NDCG@$K$ which show the same trend. We have the following observations:
\begin{itemize}[leftmargin=*]
    % \item SML-FC also performs well in both datasets while MLP perform poor in Adressa. This means stack layers and convolution layers play the most important role in transfer model. Compared with MLP results on Adressa, SML can model the short-term interest of new users more effectively. Meanwhile, our SML method is slightly affected by FC layers. Therefore, our transfer model is robust and works well.
    \item Regarding the design of transfer model, SML performs better than SML-CNN and SML-FC, which signifies the effectiveness of the hybrid structure with both CNN and fully connected layers. The improvement is attributed to the consideration of both dimension-wise relations and cross-dimension relations between the previous model $W_{t-1}$ and the new model $\hat{W}_t$. This finding is consistent with prior work~\cite{modeling}, which also verify the efficacy of jointly considering dimension-wise and cross-dimension relations in recommendation.
    
    %\item As shown in figure, the result of SML-N has poor performance in Adressa. Because, for our sequential training method, optimizing towards future performance is essential especially in some time-sensitive scenarios where one's interest drifts rapidly. Such as news recommendation or short video recommendation. 
    \item Regarding the sequential training process, the performance of SML-N is worse than SML by 18.81\% and 34.53\% on average, which validate the advantages of optimizing towards future performance. 
    Existing study on meta-learning~\cite{finn2017MAML,lambdaOpt} also demonstrated the effectiveness of optimizing model parameters towards testing (validation) data. As such, it is promising to solve periodical model retraining as a sequential meta-learning task.
    
    %\item SML-S shows that if we stop training transfer in test periods, we can still achieve satisfying performance. This supports that transfer model learn the generalization knowledge of datasets. Meanwhile, it's also benefit from updating the transfer model over time.
    %\item As we mentioned earlier, learning $\hat{w_{t}}$ alone may cause the scale not suitable for transfer. SML-FP has poor performance in Adressa, because when more and more new users appears, the scale of $\hat{w_{t}}$ will change a lot 
    \item When the transfer network is not updated during testing, the performance of SML (i.e., SML-S) drops by 7.87\% and 9.43\%. This might be caused by the drift of user interests. The performance drop signifies the importance of model retraining, which further suggests a future direction to explore online recommender updates. Lastly, SML-FP fails to achieve a comparable performance as SML on both datasets, which justifies our design of learning new model $\hat{W_{t}}$ based on the transfer output. 
    %which is might because of the unsuitable of transfer input (\ie $\hat{w_{t}}$) directly trained based on recommendation loss.
\end{itemize}

%(2)As shown in figure, the result of SML-N has poor performance in Adressa. Because, for our sequential training method, optimizing towards future performance is essential especially in some time-sensitive scenarios where one's interest drifts rapidly. Such as news recommendation or short video recommendation. 

%(3)SML-S shows that if we stop training transfer in test periods, we can still achieve satisfying performance. This supports that transfer model learn the generalization knowledge of datasets. Meanwhile, it's also benefit from updating the transfer model over time.

%(4)As we mentioned earlier, learning $\hat{w_{t}}$ alone may cause the scale not suitable for transfer. SML-FP has poor performance in Adressa, because when more and more new users appears, the scale of $\hat{w_{t}}$ will change a lot 

%In conclusion, SML-FC and MLP reflect that our transfer model is well-designed to transfer previous knowledge and the result of SML-N,SML-S and SML-FP prove that SML training process optimizes the performance on near future data.
%Figure~\ref{fig:over time}. 

%What pattern have been learned by out transfer? We show that in Figure~\ref{fig:cnn-weight-showing}.For different data, user and item transfer filter has different features. 

\subsection{Hyper-parameter Studies (RQ3)}
\begin{figure}
\centering
% \subfigure[]{
% \begin{minipage}[t]{0.2\textwidth}
% \includegraphics[width=4.0cm]{picture/hyper.pdf}
% %\caption{fig1}
% \end{minipage}%
% }%
% \subfigure[]{
% \begin{minipage}[t]{0.2\textwidth}
% \includegraphics[width=4.0cm]{picture/hyper1.pdf}
% %\caption{fig2}
% \end{minipage}%
% }%
    \mbox{
		%\hspace{-0.1in}
		\subfigure[Impact of filter number]{
		    \includegraphics[width=0.24\textwidth]{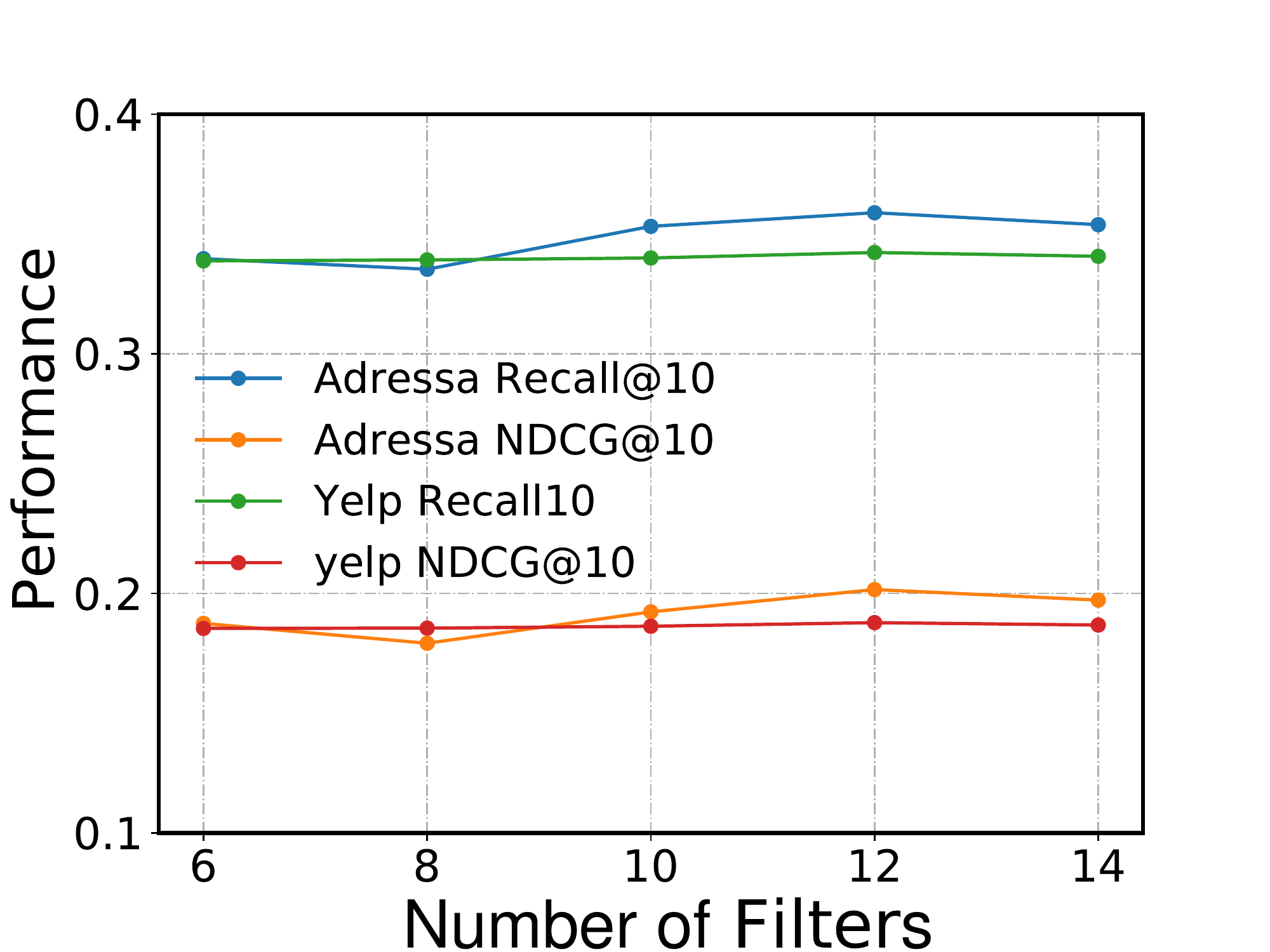} 
        }
        \hspace{-0.2in}
		\subfigure[Impact of convolution layers]{
		    \includegraphics[width=0.24\textwidth]{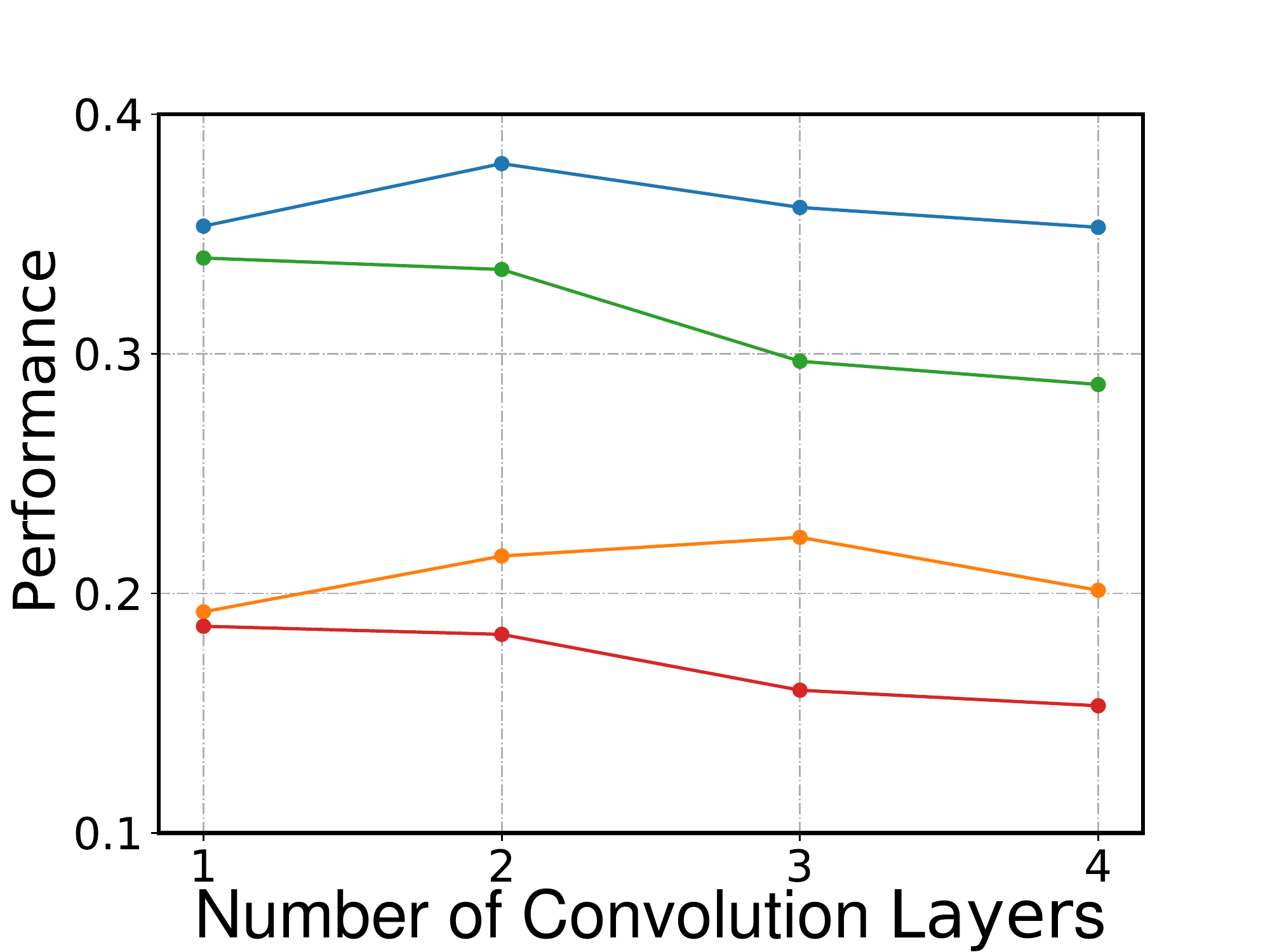} 
        }
	}
%\label{fig:hyper}
% \caption{hyper-parameters studies.(a) The influence of the number of filters in transfer. (b)the influence of the number of CNN layers in transfer. }
\vspace{-0.5cm}
\caption{Performance of SML \wrt different numbers of CNN filters and CNN layers in the transfer component.} \vspace{-8pt}
\label{fig:hypers}
\end{figure}

%The number of kernels effects the ability of  representation and stability of the model. For comparison, we use a single-layer CNN model, and adjust the kernels number from [6,8,10,12,14] without change other parameter. The result is shown in Fig8. For both dataset, performance is stable, especially Yelp. We can observe that as the number of kernels increases, the performance of Adressa is also continues to rise. 

%We fix the first CNN layer with 10 kernels and other layers with 5 kernels without changing other hyper-parameters. As the Figure7 shows, deeper CNN layers may achieve better performance ,but also reduce model stability. Therefore, we recommend to set the number of CNN layers to 1 at the beginning.

%\sudbjussuting other bsepactramets . ei uthy raper-on{Effect of Layer Numbers}
%The transfer is the component of SML to compose the previous model and the model trained on new data. We study how the key hyperparameters: \textit{the number of CNN filters} and \textit{number of CNN layers} affect the performance of SML. 
We study how the CNN architecture affects the performance, more specifically, the number of filters and the number of CNN layers. 

\subsubsection{Number of CNN Filters}
%The number of filters effects the ability of  representation and stability of the model. For comparison, 
We fix the number of convolution layer to be one and adjust the number of filters from [6,8,10,12,14]. As shown in Figure~\ref{fig:hypers}(a), the performance on Yelp is rather stable across different numbers of filters. While on Adressa, SML with more than 10 filters performs better than the filter number of 6 and 8. 
%These results suggest setting the number of filters with value between 10 and 15. 
These results suggest that for applications like Adressa where the timeliness is strong, there may exist complex relations between user short-term and long-term interests. In this case, using a large number of filters is beneficial. 

\subsubsection{Number of Convolution Layers}
We fix the first convolution layer with 10 filters and test the effect of stacking more convolution layers with 5 filters. As shown in Figure~\ref{fig:hypers}, SML achieves the best performance on Yelp with 1 convolution layer, and stacking more convolution layers degrades the performance because of overfitting. 
For Adressa, the best performance is achieved when the number of convolution layers is 2 or 3, and further increasing it also degrades the performance. 
These results suggest that the optimal number of convolution layers varies, depending on the features of the dataset. 
%characteristics
%while similar issue occurs when the transfer is deep (\ie 3 or 4 layers), slight performance improvement is achieved when increasing the layer from 1 to 2. Again, this result suggests using a more complex transfer component for applications with strong short-term user preference. Overall, we recommend to set the number of CNN layers to 1.

\subsection{In-depth Analyses (RQ4)}
%In this section, we will indirectly verify the model power to long-term and short-term interest and show why our model why can outperform the FR.At the last we will verify that our model can speed-up.
%\textbf{long-term and short-term interst.}
We conduct in-depth analyses to understand where the improvements come from compared with the Full-retrain, and scrutinize the CNN filters to interpret their rationality. 

\subsubsection{Performance of Different Interaction Types}
\begin{figure} 
	\centering

    \mbox{
		%\hspace{-0.1in}
		\subfigure[Two types of new items]{
		    \includegraphics[width=0.25\textwidth]{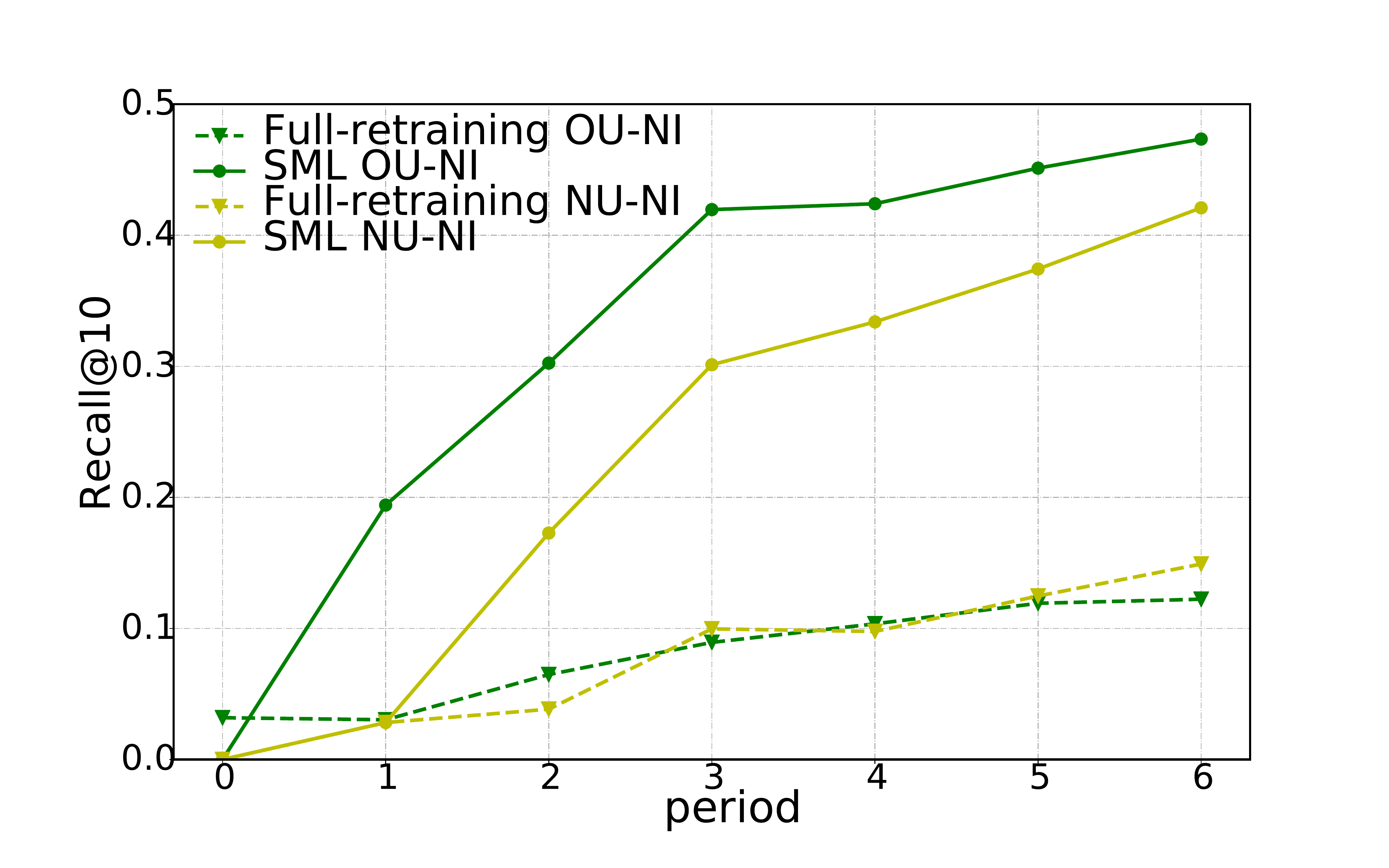}
        }
        \hspace{-0.2in}
		\subfigure[Two types of old items]{
		    \includegraphics[width=0.25\textwidth]{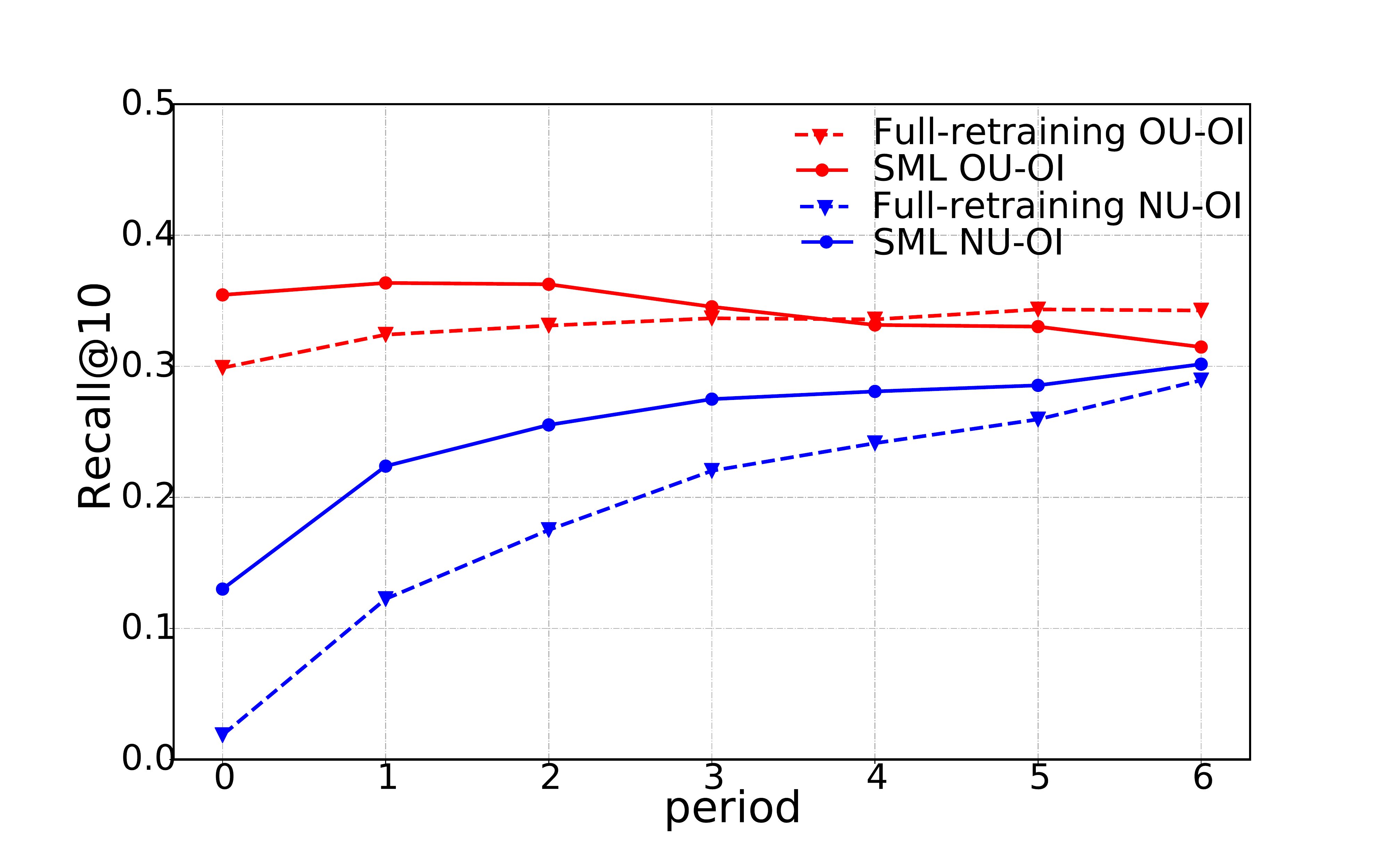}
        }
	}
	\vspace{-0.5cm}
	\caption{Recommendation performance of SML and Full-retrain on Yelp grouped by four interaction types: old user-new item (OU-NI), new user-new item (NU-NI), old user-old item (OU-OI), and new user-old item (NU-OI).}\vspace{-8pt}
	\Description{different type interaction}
	\label{fig:type perform} 
\end{figure}

%To verify whether SML has keep the long-term interests and verify the reason that why our model can outperform the Full-retrain, we design a analyze experiments on Yelp to answer these questions. 
%To further invest whether SML properly accounts for both the long-term and short-term interests,
We divide users into two groups: \textit{new users} mean the users that only occur in the testing data, otherwise \textit{old users}; same for the item side. We then cross user groups and item groups, dividing the interaction into four types: \textit{old user-new item} (OU-NI), \textit{new user-new item} (NU-NI), \textit{old user-old item} (OU-OI), and \textit{new user-old item} (NU-OI).
We then perform evaluation on each type of interactions. 
Figure \ref{fig:type perform} shows the performance of SML and Full-retrain at each testing period on the Yelp data. 
From the left subfigure (a), we observe that SML outperforms Full-retrain on the two types of new items (OU-NI and NU-NI) by a large margin. 
From the right subfigure (b), we observe that SML improves over Full-retrain on the type of new user-old item (NU-OI), while achieves a performance comparable with Full-retrain on the type of old user-old item (OU-OI). 

From these results, we draw the conclusion that the improvements of SML over Full-retrain are mainly from the recommendations for new users and new items. This shows the strong ability of SML in quickly adapting to new data that are more reflective of user short-term interests. Moreover, the performance on the interaction type of old user-old item is not degraded, which verifies the  effectiveness of SML in capturing long-term interests.

%\vspace{-0.1cm}
\subsubsection{Visualization of CNN Filters} 
\begin{figure}[t]
    \centering
    \includegraphics[width=0.48\textwidth]{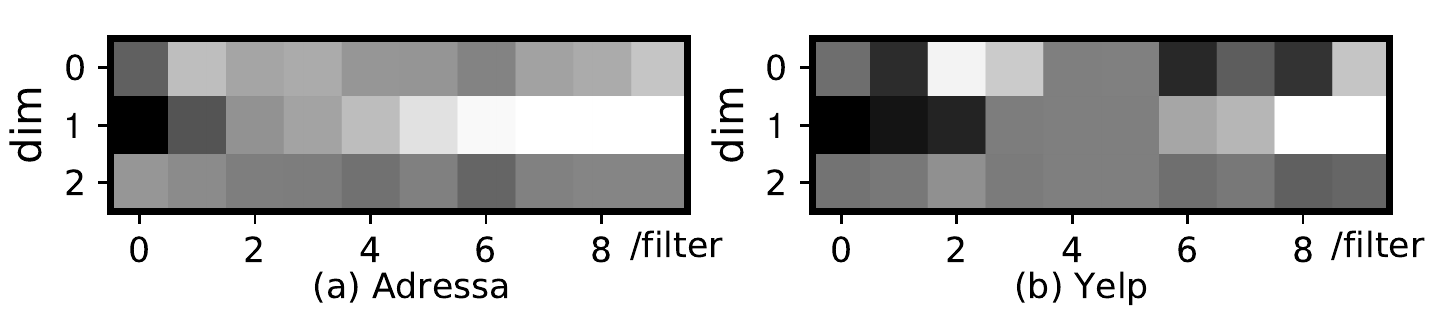}
    \vspace{-0.8cm}
    \caption{Visualization of the learned filters.}\vspace{-0.3cm}
    \label{fig:cnn-weight-showing}
\end{figure}

We study the learned CNN filters to disentangle how the transfer fuses $W_{t-1}$ and $\hat{W}_t$. In Figure~\ref{fig:cnn-weight-showing}, we visualize the learned filters of the first CNN layer in the item transfer network. Note that similar patterns can be observed in the user transfer network, which are omitted for space. 
We can see that the filters learned from Yelp and Adressa encode different patterns. 
For Adressa, the row vector corresponding to $\hat{W_t}$ (\ie $dim=1$) has higher values than the other two vectors in general, justifying that the transfer network pays more attention to recent data. 
For Yelp, we can see two special filters (the $2^{nd}$ and $8^{th}$ one), where the values at $dim=0$ (corresponding to $W_{t-1}$) and $dim=1$ have opposite sign. It means that the transfer learns to utilize the difference between $\hat{W_t}$ and $W_{t-1}$, which is beneficial to capture how user interests evolve.

\section{Related Works}\label{sec:related}
%To our knowledge, there is no existing work focusing on periodical recommender retraining. 
%In this section, we briefly review two lines of relevant work: recommendation on sequential data and meta-learning. 
\subsection{Recommendation on Sequential Data}
%By virtue of various neural networks and attention method, recommendations on sequential data make great progress.A number of recent studies on sequential recommendation are based on recurrent neural network(RNN). DREAM \cite{DREAM} makes all the user's historical interactions embedded into a RNN structure to predict the score toward items. \cite{TDSSM} captures user long-term interest using RNN and design a method to combine short-term interest and long-term interest. 
The user-item interaction data naturally forms a sequence because each interaction is associated with timestamp information. A large body of work has modeled a sequence of interactions to predict the next interaction, called as sequential~\cite{sequenceaware}, next-item/basket~\cite{NextItNet,PIF} 
 or session-based recommendation~\cite{GRU4Rec,Future}. An early representative method is Factorized Personalized Markov Chain (FPMC)~\cite{FPMC}, which models the transition between an interacted item and the previously interacted item with matrix factorization. Later work has extended the first-order modeling~\cite{TransRec}
 to high-order modeling~\cite{HOT}.
 Recently, many neural network models have been developed, wherein recurrent neural network (RNN) is a natural choice for sequence modeling~\cite{GRU4Rec}. The latest work \cite{PIF} points out a limitation of RNN that it fails to learn the personalized item frequency information in next-basket recommendation. 
 In addition, CNN has also been used for sequential recommendation~\cite{Caser,NextItNet,Future}, which is stronger than RNN in modeling the inter-independency between items. Here we consider the sequential nature of interaction data in an orthogonal way --- the model needs be periodically retrained to adapt to new data. We propose a general sequential training paradigm, which is technically viable to deploy on the sequential recommendation models. 
 
Another line of work is online or streaming recommendation~\cite{fastMF,streaming,Rstreaming}, which aims to refresh recommendations based on real-time user interactions. Several strategies have been proposed, which make different trade-offs between model freshness and accuracy. For example, \cite{fastMF,dynamicMF,EAR} perform local model updates for each new interaction, which is easy to suffer from forgetting long-term preference when many new interactions are updated. 
\cite{real-time,SIGIR18-YinHongzhi} address the issue by sampling a faction of historical interactions and mixing them with new interactions for model updating. The sampler is designed heuristically and needs be adapted manually for different recommendation domains. 
Our method does not use historical data for model refreshing, achieving good trade-off between long-term and short-term modeling by optimizing for the future performance. 

\begin{figure}
    \centering
    \includegraphics[width=0.48\textwidth]{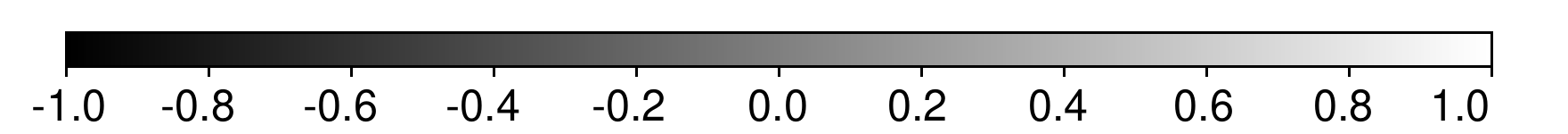}
    \vspace{-0.8cm}
\end{figure}

\subsection{Meta-Learning}
Meta-learning, or learning to learn, aims to quickly and effectively adapt to new tasks by using the prior experience learned from the related tasks~\cite{finn2017MAML,meta-learner}. A representative method is Model-Agnostic Meta-Learning (MAML)~\cite{finn2017MAML}, which represents a general paradigm that uses the testing data of a task to optimize the training process (e.g., initialization and hyper-parameters~\cite{bilevel}). %ICML2018 Bilevel Programming for Hyperparameter Optimization and Meta-Learning
The idea of MAML has been taken to solve the cold-start recommendation problem~\cite{warmup,s2meta,LWA,MeLU,MetaCS}. For example, by treating each user as a task, \cite{MeLU,MetaCS} learn how to generalize well with few iterations. Besides, some works have utilized meta learning to select recommendation algorithms \cite{meta-selectS} and a recent work \cite{lambdaOpt} proposes $\lambda$Opt to optimize regularization hyper-parameters based on the validation performance. Inspired by MAML and $\lambda$Opt, we optimize the model retraining process based on the future validation data, proposing a new training paradigm for sequential user-item interactions. 

%In addition to the above works, lifelong learning, such as \cite{ewc,ewc2}, is also relevant to our work. However, the goal of lifelong learning is to make model not catastrophically forget old tasks when learning new task, and time-order of tasks is not considered, it's different from retraining setting for recommender system.
Beyond recommendation, lifelong learning~\cite{ewc,ewc2} is weakly relevant to this work, which aims to learn different tasks in a sequence. Several strategies were devised to avoid catastrophically forget old tasks when learning a new task so that all tasks are well served. For instance, \cite{ewc} remembers old tasks by selectively slowing down learning on the weights important for those tasks. As the key objective of lifelong learning is to serve the current task (on $D_t$) without suffering performance on previous task, these strategies are not suitable for the sequential training of recommendation model where optimizing the model for better serving the future task (on $D_{t+1}$) is of importance.

\section{conclusion}\label{sec:conclude}
%With the belief that the previous model has contained all knowledge of previous data, We design a retraining methods(SML) by solving retraining problem as a optimizable problem with only current data. 
In this work, we investigated the retraining of recommender models. 
We formulated the task of recommender retraining, which aims to achieve good generalization on next-period data by modeling the newly collected data. 
To address the task, we proposed a sequential meta-learning (SML) approach, which consists of 1) an expressive transfer network that converts the previous model to a new model based on the newly collected data, and 2) a sequential training method that effectively utilizes the next-period data to learn the transfer network. We conducted experiments on two datasets of different properties, providing extensive results and analyses on the effectiveness, efficiency, and rationality of SML. 

In future, we will extend our generic training paradigm to a wide range of recommender models, such as the recently emerging graph neural networks~\cite{ngcf,LightGCN} that are more effective for collaborative filtering, and factorization machines~\cite{FM,NFM} that can incorporate various side information and sequential recommender models~\cite{Caser}. Currently, we do not consider online learning strategy (e.g., bandit methods and local parameter updates) in each serving period, and we plan to take it into account. Lastly, we will develop personalized meta-learning mechanisms that optimize the learning process for different users differently. 
%In the future, we will verify our methods in network-based methods include deep CF models such as NCF\cite{NCF}, the model that can process side information,such as NFM\cite{NFM},and sequential models,such as Caser\cite{Caser}. In addition, our model doesn't consider real-time updating in online, solving retraining(updating) in online is another direction for future. Besides, about the transfer component in SML for CF model, all the users/items share the same transfer. In the future, a transfer with personalization which can provide different type "transfer knowledge" solution for different type users/items may be improve SML performance.
%\vspace{-0.5cm}
% \begin{acks}
% This work is supported by the National Natural Science Foundation of China (61972372, U19A2079,61725203).
% \end{acks}
% \vspace{-0.2cm}
\bibliographystyle{ACM-Reference-Format}
\bibliography{sample-base}


\begin{thebibliography}{47}

%%% the Web address.

\ifx \showCODEN    \undefined \def \showCODEN     #1{\unskip}     \fi
\ifx \showDOI      \undefined \def \showDOI       #1{#1}\fi
\ifx \showISBNx    \undefined \def \showISBNx     #1{\unskip}     \fi
\ifx \showISBNxiii \undefined \def \showISBNxiii  #1{\unskip}     \fi
\ifx \showISSN     \undefined \def \showISSN      #1{\unskip}     \fi
\ifx \showLCCN     \undefined \def \showLCCN      #1{\unskip}     \fi
\ifx \shownote     \undefined \def \shownote      #1{#1}          \fi
\ifx \showarticletitle \undefined \def \showarticletitle #1{#1}   \fi
\ifx \showURL      \undefined \def \showURL       {\relax}        \fi
% invisible to TeX
\providecommand\bibfield[2]{#2}
\providecommand\bibinfo[2]{#2}
\providecommand\natexlab[1]{#1}
\providecommand\showeprint[2][]{arXiv:#2}

\bibitem[\protect\citeauthoryear{Beutel, Covington, Jain, Xu, Li, Gatto, and
  Chi}{Beutel et~al\mbox{.}}{2018}]%
        {latent-cross}
\bibfield{author}{\bibinfo{person}{Alex Beutel}, \bibinfo{person}{Paul
  Covington}, \bibinfo{person}{Sagar Jain}, \bibinfo{person}{Can Xu},
  \bibinfo{person}{Jia Li}, \bibinfo{person}{Vince Gatto}, {and}
  \bibinfo{person}{Ed~H Chi}.} \bibinfo{year}{2018}\natexlab{}.
\newblock \showarticletitle{Latent cross: Making use of context in recurrent
  recommender systems}. In \bibinfo{booktitle}{\emph{WSDM}}.
  \bibinfo{pages}{46--54}.
\newblock


\bibitem[\protect\citeauthoryear{Bharadhwaj}{Bharadhwaj}{2019}]%
        {MetaCS}
\bibfield{author}{\bibinfo{person}{Homanga Bharadhwaj}.}
  \bibinfo{year}{2019}\natexlab{}.
\newblock \showarticletitle{Meta-Learning for User Cold-Start Recommendation}.
  In \bibinfo{booktitle}{\emph{IJCNN}}. \bibinfo{pages}{1--8}.
\newblock


\bibitem[\protect\citeauthoryear{Chang, Zhang, Tang, Yin, Chang,
  Hasegawa-Johnson, and Huang}{Chang et~al\mbox{.}}{2017}]%
        {streaming}
\bibfield{author}{\bibinfo{person}{Shiyu Chang}, \bibinfo{person}{Yang Zhang},
  \bibinfo{person}{Jiliang Tang}, \bibinfo{person}{Dawei Yin},
  \bibinfo{person}{Yi Chang}, \bibinfo{person}{Mark~A Hasegawa-Johnson}, {and}
  \bibinfo{person}{Thomas~S Huang}.} \bibinfo{year}{2017}\natexlab{}.
\newblock \showarticletitle{Streaming recommender systems}. In
  \bibinfo{booktitle}{\emph{WWW}}. \bibinfo{pages}{525--534}.
\newblock


\bibitem[\protect\citeauthoryear{Chen, Chen, He, Gao, Li, Lou, and Wang}{Chen
  et~al\mbox{.}}{2019}]%
        {lambdaOpt}
\bibfield{author}{\bibinfo{person}{Yihong Chen}, \bibinfo{person}{Bei Chen},
  \bibinfo{person}{Xiangnan He}, \bibinfo{person}{Chen Gao},
  \bibinfo{person}{Yong Li}, \bibinfo{person}{Jian{-}Guang Lou}, {and}
  \bibinfo{person}{Yue Wang}.} \bibinfo{year}{2019}\natexlab{}.
\newblock \showarticletitle{{\(\lambda\)}Opt: Learn to Regularize Recommender
  Models in Finer Levels}. In \bibinfo{booktitle}{\emph{SIGKDD}}.
  \bibinfo{pages}{978--986}.
\newblock


\bibitem[\protect\citeauthoryear{Devooght, Kourtellis, and Mantrach}{Devooght
  et~al\mbox{.}}{2015}]%
        {dynamicMF}
\bibfield{author}{\bibinfo{person}{Robin Devooght}, \bibinfo{person}{Nicolas
  Kourtellis}, {and} \bibinfo{person}{Amin Mantrach}.}
  \bibinfo{year}{2015}\natexlab{}.
\newblock \showarticletitle{Dynamic matrix factorization with priors on unknown
  values}. In \bibinfo{booktitle}{\emph{SIGKDD}}. \bibinfo{pages}{189--198}.
\newblock


\bibitem[\protect\citeauthoryear{Diaz-Aviles, Drumond, Schmidt-Thieme, and
  Nejdl}{Diaz-Aviles et~al\mbox{.}}{2012}]%
        {real-time}
\bibfield{author}{\bibinfo{person}{Ernesto Diaz-Aviles}, \bibinfo{person}{Lucas
  Drumond}, \bibinfo{person}{Lars Schmidt-Thieme}, {and}
  \bibinfo{person}{Wolfgang Nejdl}.} \bibinfo{year}{2012}\natexlab{}.
\newblock \showarticletitle{Real-time top-n recommendation in social streams}.
  In \bibinfo{booktitle}{\emph{RecSys}}. \bibinfo{pages}{59--66}.
\newblock


\bibitem[\protect\citeauthoryear{Du, He, Yuan, Tang, Qin, and Chua}{Du
  et~al\mbox{.}}{2019a}]%
        {modeling}
\bibfield{author}{\bibinfo{person}{Xiaoyu Du}, \bibinfo{person}{Xiangnan He},
  \bibinfo{person}{Fajie Yuan}, \bibinfo{person}{Jinhui Tang},
  \bibinfo{person}{Zhiguang Qin}, {and} \bibinfo{person}{Tat-Seng Chua}.}
  \bibinfo{year}{2019}\natexlab{a}.
\newblock \showarticletitle{Modeling Embedding Dimension Correlations via
  Convolutional Neural Collaborative Filtering}.
\newblock \bibinfo{journal}{\emph{TOIS}} \bibinfo{volume}{37},
  \bibinfo{number}{4} (\bibinfo{year}{2019}), \bibinfo{pages}{47:1--47:22}.
\newblock


\bibitem[\protect\citeauthoryear{Du, Wang, Yang, Zhou, and Tang}{Du
  et~al\mbox{.}}{2019b}]%
        {s2meta}
\bibfield{author}{\bibinfo{person}{Zhengxiao Du}, \bibinfo{person}{Xiaowei
  Wang}, \bibinfo{person}{Hongxia Yang}, \bibinfo{person}{Jingren Zhou}, {and}
  \bibinfo{person}{Jie Tang}.} \bibinfo{year}{2019}\natexlab{b}.
\newblock \showarticletitle{Sequential Scenario-Specific Meta Learner for
  Online Recommendation}. In \bibinfo{booktitle}{\emph{SIGKDD}}.
  \bibinfo{pages}{2895--2904}.
\newblock


\bibitem[\protect\citeauthoryear{Finn, Abbeel, and Levine}{Finn
  et~al\mbox{.}}{2017}]%
        {finn2017MAML}
\bibfield{author}{\bibinfo{person}{Chelsea Finn}, \bibinfo{person}{Pieter
  Abbeel}, {and} \bibinfo{person}{Sergey Levine}.}
  \bibinfo{year}{2017}\natexlab{}.
\newblock \showarticletitle{Model-agnostic meta-learning for fast adaptation of
  deep networks}. In \bibinfo{booktitle}{\emph{ICML}},
  Vol.~\bibinfo{volume}{70}. \bibinfo{pages}{1126--1135}.
\newblock


\bibitem[\protect\citeauthoryear{Franceschi, Frasconi, Salzo, Grazzi, and
  Pontil}{Franceschi et~al\mbox{.}}{2018}]%
        {bilevel}
\bibfield{author}{\bibinfo{person}{Luca Franceschi}, \bibinfo{person}{Paolo
  Frasconi}, \bibinfo{person}{Saverio Salzo}, \bibinfo{person}{Riccardo
  Grazzi}, {and} \bibinfo{person}{Massimiliano Pontil}.}
  \bibinfo{year}{2018}\natexlab{}.
\newblock \showarticletitle{Bilevel Programming for Hyperparameter Optimization
  and Meta-Learning}. In \bibinfo{booktitle}{\emph{ICML}},
  Vol.~\bibinfo{volume}{80}. \bibinfo{pages}{1563--1572}.
\newblock


\bibitem[\protect\citeauthoryear{Gulla, Zhang, Liu, {\"O}zg{\"o}bek, and
  Su}{Gulla et~al\mbox{.}}{2017}]%
        {adressa}
\bibfield{author}{\bibinfo{person}{Jon~Atle Gulla}, \bibinfo{person}{Lemei
  Zhang}, \bibinfo{person}{Peng Liu}, \bibinfo{person}{{\"O}zlem
  {\"O}zg{\"o}bek}, {and} \bibinfo{person}{Xiaomeng Su}.}
  \bibinfo{year}{2017}\natexlab{}.
\newblock \showarticletitle{The Adressa dataset for news recommendation}. In
  \bibinfo{booktitle}{\emph{WI}}. \bibinfo{pages}{1042--1048}.
\newblock


\bibitem[\protect\citeauthoryear{He, Kang, and McAuley}{He
  et~al\mbox{.}}{2017a}]%
        {TransRec}
\bibfield{author}{\bibinfo{person}{Ruining He}, \bibinfo{person}{Wang-Cheng
  Kang}, {and} \bibinfo{person}{Julian McAuley}.}
  \bibinfo{year}{2017}\natexlab{a}.
\newblock \showarticletitle{Translation-based recommendation}. In
  \bibinfo{booktitle}{\emph{RecSys}}. \bibinfo{pages}{161--169}.
\newblock


\bibitem[\protect\citeauthoryear{He and Chua}{He and Chua}{2017}]%
        {NFM}
\bibfield{author}{\bibinfo{person}{Xiangnan He} {and} \bibinfo{person}{Tat-Seng
  Chua}.} \bibinfo{year}{2017}\natexlab{}.
\newblock \showarticletitle{Neural factorization machines for sparse predictive
  analytics}. In \bibinfo{booktitle}{\emph{SIGIR}}. \bibinfo{pages}{355--364}.
\newblock


\bibitem[\protect\citeauthoryear{He, Deng, Wang, Li, Zhang, and Wang}{He
  et~al\mbox{.}}{2020}]%
        {LightGCN}
\bibfield{author}{\bibinfo{person}{Xiangnan He}, \bibinfo{person}{Kuan Deng},
  \bibinfo{person}{Xiang Wang}, \bibinfo{person}{Yan Li},
  \bibinfo{person}{Yongdong Zhang}, {and} \bibinfo{person}{Meng Wang}.}
  \bibinfo{year}{2020}\natexlab{}.
\newblock \showarticletitle{LightGCN: Simplifying and Powering Graph
  Convolution Network for Recommendation}. In
  \bibinfo{booktitle}{\emph{SIGIR}}.
\newblock


\bibitem[\protect\citeauthoryear{He, Liao, Zhang, Nie, Hu, and Chua}{He
  et~al\mbox{.}}{2017b}]%
        {NCF}
\bibfield{author}{\bibinfo{person}{Xiangnan He}, \bibinfo{person}{Lizi Liao},
  \bibinfo{person}{Hanwang Zhang}, \bibinfo{person}{Liqiang Nie},
  \bibinfo{person}{Xia Hu}, {and} \bibinfo{person}{Tat-Seng Chua}.}
  \bibinfo{year}{2017}\natexlab{b}.
\newblock \showarticletitle{Neural collaborative filtering}. In
  \bibinfo{booktitle}{\emph{WWW}}. \bibinfo{pages}{173--182}.
\newblock


\bibitem[\protect\citeauthoryear{He, Zhang, Kan, and Chua}{He
  et~al\mbox{.}}{2016}]%
        {fastMF}
\bibfield{author}{\bibinfo{person}{Xiangnan He}, \bibinfo{person}{Hanwang
  Zhang}, \bibinfo{person}{Min{-}Yen Kan}, {and} \bibinfo{person}{Tat{-}Seng
  Chua}.} \bibinfo{year}{2016}\natexlab{}.
\newblock \showarticletitle{Fast Matrix Factorization for Online Recommendation
  with Implicit Feedback}. In \bibinfo{booktitle}{\emph{SIGIR}}.
  \bibinfo{pages}{549--558}.
\newblock


\bibitem[\protect\citeauthoryear{Hendrycks and Gimpel}{Hendrycks and
  Gimpel}{2016}]%
        {GELU}
\bibfield{author}{\bibinfo{person}{Dan Hendrycks} {and} \bibinfo{person}{Kevin
  Gimpel}.} \bibinfo{year}{2016}\natexlab{}.
\newblock \showarticletitle{Bridging Nonlinearities and Stochastic Regularizers
  with Gaussian Error Linear Units}.
\newblock \bibinfo{journal}{\emph{CoRR}}  \bibinfo{volume}{abs/1606.08415}
  (\bibinfo{year}{2016}).
\newblock


\bibitem[\protect\citeauthoryear{Hidasi, Karatzoglou, Baltrunas, and
  Tikk}{Hidasi et~al\mbox{.}}{2016}]%
        {GRU4Rec}
\bibfield{author}{\bibinfo{person}{Bal{\'{a}}zs Hidasi},
  \bibinfo{person}{Alexandros Karatzoglou}, \bibinfo{person}{Linas Baltrunas},
  {and} \bibinfo{person}{Domonkos Tikk}.} \bibinfo{year}{2016}\natexlab{}.
\newblock \showarticletitle{Session-based Recommendations with Recurrent Neural
  Networks}. In \bibinfo{booktitle}{\emph{ICLR}}.
\newblock


\bibitem[\protect\citeauthoryear{Hornik}{Hornik}{1991}]%
        {mlpapproximation}
\bibfield{author}{\bibinfo{person}{Kurt Hornik}.}
  \bibinfo{year}{1991}\natexlab{}.
\newblock \showarticletitle{Approximation capabilities of multilayer
  feedforward networks}.
\newblock \bibinfo{journal}{\emph{Neural networks}} \bibinfo{volume}{4},
  \bibinfo{number}{2} (\bibinfo{year}{1991}), \bibinfo{pages}{251--257}.
\newblock


\bibitem[\protect\citeauthoryear{Hu, He, Gao, and Zhang}{Hu
  et~al\mbox{.}}{2020}]%
        {PIF}
\bibfield{author}{\bibinfo{person}{Haoji Hu}, \bibinfo{person}{Xiangnan He},
  \bibinfo{person}{Jinyang Gao}, {and} \bibinfo{person}{Zhi-Li Zhang}.}
  \bibinfo{year}{2020}\natexlab{}.
\newblock \showarticletitle{Modeling Personalized Item Frequency Information
  for Next-basket Recommendation}. In \bibinfo{booktitle}{\emph{SIGIR}}.
\newblock


\bibitem[\protect\citeauthoryear{Jamal and Qi}{Jamal and Qi}{2019}]%
        {TAML}
\bibfield{author}{\bibinfo{person}{Muhammad~Abdullah Jamal} {and}
  \bibinfo{person}{Guo-Jun Qi}.} \bibinfo{year}{2019}\natexlab{}.
\newblock \showarticletitle{Task Agnostic Meta-Learning for Few-Shot Learning}.
  In \bibinfo{booktitle}{\emph{CVPR}}. \bibinfo{pages}{11719--11727}.
\newblock


\bibitem[\protect\citeauthoryear{Kingma and Ba}{Kingma and Ba}{2015}]%
        {adam}
\bibfield{author}{\bibinfo{person}{Diederik~P. Kingma} {and}
  \bibinfo{person}{Jimmy Ba}.} \bibinfo{year}{2015}\natexlab{}.
\newblock \showarticletitle{Adam: {A} Method for Stochastic Optimization}. In
  \bibinfo{booktitle}{\emph{ICLR}}.
\newblock


\bibitem[\protect\citeauthoryear{Kirkpatrick, Pascanu, Rabinowitz, Veness,
  Desjardins, Rusu, Milan, Quan, Ramalho, Grabskabarwinska,
  et~al\mbox{.}}{Kirkpatrick et~al\mbox{.}}{2017}]%
        {ewc}
\bibfield{author}{\bibinfo{person}{James Kirkpatrick}, \bibinfo{person}{Razvan
  Pascanu}, \bibinfo{person}{Neil~C Rabinowitz}, \bibinfo{person}{Joel Veness},
  \bibinfo{person}{Guillaume Desjardins}, \bibinfo{person}{Andrei~A Rusu},
  \bibinfo{person}{Kieran Milan}, \bibinfo{person}{John Quan},
  \bibinfo{person}{Tiago Ramalho}, \bibinfo{person}{Agnieszka
  Grabskabarwinska}, {et~al\mbox{.}}} \bibinfo{year}{2017}\natexlab{}.
\newblock \showarticletitle{Overcoming catastrophic forgetting in neural
  networks}.
\newblock \bibinfo{journal}{\emph{PNAS}} \bibinfo{volume}{114},
  \bibinfo{number}{13} (\bibinfo{year}{2017}), \bibinfo{pages}{3521--3526}.
\newblock


\bibitem[\protect\citeauthoryear{Lee, Im, Jang, Cho, and Chung}{Lee
  et~al\mbox{.}}{2019}]%
        {MeLU}
\bibfield{author}{\bibinfo{person}{Hoyeop Lee}, \bibinfo{person}{Jinbae Im},
  \bibinfo{person}{Seongwon Jang}, \bibinfo{person}{Hyunsouk Cho}, {and}
  \bibinfo{person}{Sehee Chung}.} \bibinfo{year}{2019}\natexlab{}.
\newblock \showarticletitle{MeLU: Meta-Learned User Preference Estimator for
  Cold-Start Recommendation}. In \bibinfo{booktitle}{\emph{SIGKDD}}.
  \bibinfo{pages}{1073--1082}.
\newblock


\bibitem[\protect\citeauthoryear{Lei, He, Miao, Wu, Hong, Kan, and Chua}{Lei
  et~al\mbox{.}}{2020}]%
        {EAR}
\bibfield{author}{\bibinfo{person}{Wenqiang Lei}, \bibinfo{person}{Xiangnan
  He}, \bibinfo{person}{Yisong Miao}, \bibinfo{person}{Qingyun Wu},
  \bibinfo{person}{Richang Hong}, \bibinfo{person}{Min{-}Yen Kan}, {and}
  \bibinfo{person}{Tat{-}Seng Chua}.} \bibinfo{year}{2020}\natexlab{}.
\newblock \showarticletitle{Estimation-Action-Reflection: Towards Deep
  Interaction Between Conversational and Recommender Systems}. In
  \bibinfo{booktitle}{\emph{WSDM}}. \bibinfo{pages}{304--312}.
\newblock


\bibitem[\protect\citeauthoryear{Lian, Zhou, Zhang, Chen, Xie, and Sun}{Lian
  et~al\mbox{.}}{2018}]%
        {xDeepFM}
\bibfield{author}{\bibinfo{person}{Jianxun Lian}, \bibinfo{person}{Xiaohuan
  Zhou}, \bibinfo{person}{Fuzheng Zhang}, \bibinfo{person}{Zhongxia Chen},
  \bibinfo{person}{Xing Xie}, {and} \bibinfo{person}{Guangzhong Sun}.}
  \bibinfo{year}{2018}\natexlab{}.
\newblock \showarticletitle{xdeepfm: Combining explicit and implicit feature
  interactions for recommender systems}. In \bibinfo{booktitle}{\emph{SIGKDD}}.
  \bibinfo{pages}{1754--1763}.
\newblock


\bibitem[\protect\citeauthoryear{Lopez{-}Paz and Ranzato}{Lopez{-}Paz and
  Ranzato}{2017}]%
        {ewc2}
\bibfield{author}{\bibinfo{person}{David Lopez{-}Paz} {and}
  \bibinfo{person}{Marc'Aurelio Ranzato}.} \bibinfo{year}{2017}\natexlab{}.
\newblock \showarticletitle{Gradient Episodic Memory for Continual Learning}.
  In \bibinfo{booktitle}{\emph{NeurlPS 2017}}. \bibinfo{pages}{6467--6476}.
\newblock


\bibitem[\protect\citeauthoryear{Pan, Li, Ao, Tang, and He}{Pan
  et~al\mbox{.}}{2019}]%
        {warmup}
\bibfield{author}{\bibinfo{person}{Feiyang Pan}, \bibinfo{person}{Shuokai Li},
  \bibinfo{person}{Xiang Ao}, \bibinfo{person}{Pingzhong Tang}, {and}
  \bibinfo{person}{Qing He}.} \bibinfo{year}{2019}\natexlab{}.
\newblock \showarticletitle{Warm Up Cold-start Advertisements: Improving {CTR}
  Predictions via Learning to Learn {ID} Embeddings}. In
  \bibinfo{booktitle}{\emph{SIGIR}}. \bibinfo{pages}{695--704}.
\newblock


\bibitem[\protect\citeauthoryear{Quadrana, Cremonesi, and Jannach}{Quadrana
  et~al\mbox{.}}{2018}]%
        {sequenceaware}
\bibfield{author}{\bibinfo{person}{Massimo Quadrana}, \bibinfo{person}{Paolo
  Cremonesi}, {and} \bibinfo{person}{Dietmar Jannach}.}
  \bibinfo{year}{2018}\natexlab{}.
\newblock \showarticletitle{Sequence-aware recommender systems}.
\newblock \bibinfo{journal}{\emph{ACM Computing Surveys (CSUR)}}
  \bibinfo{volume}{51}, \bibinfo{number}{4} (\bibinfo{year}{2018}),
  \bibinfo{pages}{66:1--66:36}.
\newblock


\bibitem[\protect\citeauthoryear{Ravi and Larochelle}{Ravi and
  Larochelle}{2017}]%
        {meta-learner}
\bibfield{author}{\bibinfo{person}{Sachin Ravi} {and} \bibinfo{person}{Hugo
  Larochelle}.} \bibinfo{year}{2017}\natexlab{}.
\newblock \showarticletitle{Optimization as a Model for Few-Shot Learning}. In
  \bibinfo{booktitle}{\emph{ICLR}}.
\newblock


\bibitem[\protect\citeauthoryear{Ren, Chi, and Jintao}{Ren
  et~al\mbox{.}}{2019}]%
        {meta-selectS}
\bibfield{author}{\bibinfo{person}{Yi Ren}, \bibinfo{person}{Cuirong Chi},
  {and} \bibinfo{person}{Zhang Jintao}.} \bibinfo{year}{2019}\natexlab{}.
\newblock \showarticletitle{A Survey of Personalized Recommendation Algorithm
  Selection Based on Meta-learning}. In \bibinfo{booktitle}{\emph{CSIA}}.
  \bibinfo{pages}{1383--1388}.
\newblock


\bibitem[\protect\citeauthoryear{Rendle}{Rendle}{2010}]%
        {FM}
\bibfield{author}{\bibinfo{person}{Steffen Rendle}.}
  \bibinfo{year}{2010}\natexlab{}.
\newblock \showarticletitle{Factorization Machines}. In
  \bibinfo{booktitle}{\emph{ICDM}}. \bibinfo{pages}{995--1000}.
\newblock


\bibitem[\protect\citeauthoryear{Rendle, Freudenthaler, Gantner, and
  Schmidt-Thieme}{Rendle et~al\mbox{.}}{2009}]%
        {BPR}
\bibfield{author}{\bibinfo{person}{Steffen Rendle}, \bibinfo{person}{Christoph
  Freudenthaler}, \bibinfo{person}{Zeno Gantner}, {and} \bibinfo{person}{Lars
  Schmidt-Thieme}.} \bibinfo{year}{2009}\natexlab{}.
\newblock \showarticletitle{BPR: Bayesian personalized ranking from implicit
  feedback}. In \bibinfo{booktitle}{\emph{UAI}}. \bibinfo{pages}{452--461}.
\newblock


\bibitem[\protect\citeauthoryear{Rendle, Freudenthaler, and
  Schmidt-Thieme}{Rendle et~al\mbox{.}}{2010}]%
        {FPMC}
\bibfield{author}{\bibinfo{person}{Steffen Rendle}, \bibinfo{person}{Christoph
  Freudenthaler}, {and} \bibinfo{person}{Lars Schmidt-Thieme}.}
  \bibinfo{year}{2010}\natexlab{}.
\newblock \showarticletitle{Factorizing personalized markov chains for
  next-basket recommendation}. In \bibinfo{booktitle}{\emph{WWW}}.
  \bibinfo{pages}{811--820}.
\newblock


\bibitem[\protect\citeauthoryear{Rendle and Schmidt{-}Thieme}{Rendle and
  Schmidt{-}Thieme}{2008}]%
        {online-update}
\bibfield{author}{\bibinfo{person}{Steffen Rendle} {and} \bibinfo{person}{Lars
  Schmidt{-}Thieme}.} \bibinfo{year}{2008}\natexlab{}.
\newblock \showarticletitle{Online-updating regularized kernel matrix
  factorization models for large-scale recommender systems}. In
  \bibinfo{booktitle}{\emph{RecSys}}. \bibinfo{pages}{251--258}.
\newblock


\bibitem[\protect\citeauthoryear{Sculley, Holt, Golovin, Davydov, Phillips,
  Ebner, Chaudhary, Young, Crespo, and Dennison}{Sculley et~al\mbox{.}}{2015}]%
        {hidden}
\bibfield{author}{\bibinfo{person}{David Sculley}, \bibinfo{person}{Gary Holt},
  \bibinfo{person}{Daniel Golovin}, \bibinfo{person}{Eugene Davydov},
  \bibinfo{person}{Todd Phillips}, \bibinfo{person}{Dietmar Ebner},
  \bibinfo{person}{Vinay Chaudhary}, \bibinfo{person}{Michael Young},
  \bibinfo{person}{Jean-Francois Crespo}, {and} \bibinfo{person}{Dan
  Dennison}.} \bibinfo{year}{2015}\natexlab{}.
\newblock \showarticletitle{Hidden technical debt in machine learning systems}.
  In \bibinfo{booktitle}{\emph{NeurlPS}}. \bibinfo{pages}{2503--2511}.
\newblock


\bibitem[\protect\citeauthoryear{Subbian, Aggarwal, and Hegde}{Subbian
  et~al\mbox{.}}{2016}]%
        {Rstreaming}
\bibfield{author}{\bibinfo{person}{Karthik Subbian}, \bibinfo{person}{Charu
  Aggarwal}, {and} \bibinfo{person}{Kshiteesh Hegde}.}
  \bibinfo{year}{2016}\natexlab{}.
\newblock \showarticletitle{Recommendations for streaming data}. In
  \bibinfo{booktitle}{\emph{CIKM}}. \bibinfo{pages}{2185--2190}.
\newblock


\bibitem[\protect\citeauthoryear{Tang and Wang}{Tang and Wang}{2018}]%
        {Caser}
\bibfield{author}{\bibinfo{person}{Jiaxi Tang} {and} \bibinfo{person}{Ke
  Wang}.} \bibinfo{year}{2018}\natexlab{}.
\newblock \showarticletitle{Personalized top-n sequential recommendation via
  convolutional sequence embedding}. In \bibinfo{booktitle}{\emph{WSDM}}.
  \bibinfo{pages}{565--573}.
\newblock


\bibitem[\protect\citeauthoryear{Vartak, Thiagarajan, Miranda, Bratman, and
  Larochelle}{Vartak et~al\mbox{.}}{2017}]%
        {LWA}
\bibfield{author}{\bibinfo{person}{Manasi Vartak}, \bibinfo{person}{Arvind
  Thiagarajan}, \bibinfo{person}{Conrado Miranda}, \bibinfo{person}{Jeshua
  Bratman}, {and} \bibinfo{person}{Hugo Larochelle}.}
  \bibinfo{year}{2017}\natexlab{}.
\newblock \showarticletitle{A meta-learning perspective on cold-start
  recommendations for items}. In \bibinfo{booktitle}{\emph{NeurlPS}}.
  \bibinfo{pages}{6904--6914}.
\newblock


\bibitem[\protect\citeauthoryear{Vitter}{Vitter}{1985}]%
        {sampling-reservoir}
\bibfield{author}{\bibinfo{person}{Jeffrey~S Vitter}.}
  \bibinfo{year}{1985}\natexlab{}.
\newblock \showarticletitle{Random sampling with a reservoir}.
\newblock \bibinfo{journal}{\emph{TOMS}} \bibinfo{volume}{11},
  \bibinfo{number}{1}, \bibinfo{pages}{37--57}.
\newblock


\bibitem[\protect\citeauthoryear{Wang, Yin, Hu, Lian, Wang, and Huang}{Wang
  et~al\mbox{.}}{2018a}]%
        {NMRN}
\bibfield{author}{\bibinfo{person}{Qinyong Wang}, \bibinfo{person}{Hongzhi
  Yin}, \bibinfo{person}{Zhiting Hu}, \bibinfo{person}{Defu Lian},
  \bibinfo{person}{Hao Wang}, {and} \bibinfo{person}{Zi Huang}.}
  \bibinfo{year}{2018}\natexlab{a}.
\newblock \showarticletitle{Neural memory streaming recommender networks with
  adversarial training}. In \bibinfo{booktitle}{\emph{SIGKDD}}.
  \bibinfo{pages}{2467--2475}.
\newblock


\bibitem[\protect\citeauthoryear{Wang, Yin, Huang, Wang, Du, and Nguyen}{Wang
  et~al\mbox{.}}{2018b}]%
        {SIGIR18-YinHongzhi}
\bibfield{author}{\bibinfo{person}{Weiqing Wang}, \bibinfo{person}{Hongzhi
  Yin}, \bibinfo{person}{Zi Huang}, \bibinfo{person}{Qinyong Wang},
  \bibinfo{person}{Xingzhong Du}, {and} \bibinfo{person}{Quoc Viet~Hung
  Nguyen}.} \bibinfo{year}{2018}\natexlab{b}.
\newblock \showarticletitle{Streaming ranking based recommender systems}. In
  \bibinfo{booktitle}{\emph{SIGIR}}. \bibinfo{pages}{525--534}.
\newblock


\bibitem[\protect\citeauthoryear{Wang, He, Cao, Liu, and Chua}{Wang
  et~al\mbox{.}}{2019a}]%
        {KGAT}
\bibfield{author}{\bibinfo{person}{Xiang Wang}, \bibinfo{person}{Xiangnan He},
  \bibinfo{person}{Yixin Cao}, \bibinfo{person}{Meng Liu}, {and}
  \bibinfo{person}{Tat{-}Seng Chua}.} \bibinfo{year}{2019}\natexlab{a}.
\newblock \showarticletitle{{KGAT:} Knowledge Graph Attention Network for
  Recommendation}. In \bibinfo{booktitle}{\emph{SIGKDD}}.
  \bibinfo{pages}{950--958}.
\newblock


\bibitem[\protect\citeauthoryear{Wang, He, Wang, Feng, and Chua}{Wang
  et~al\mbox{.}}{2019b}]%
        {ngcf}
\bibfield{author}{\bibinfo{person}{Xiang Wang}, \bibinfo{person}{Xiangnan He},
  \bibinfo{person}{Meng Wang}, \bibinfo{person}{Fuli Feng}, {and}
  \bibinfo{person}{Tat{-}Seng Chua}.} \bibinfo{year}{2019}\natexlab{b}.
\newblock \showarticletitle{Neural Graph Collaborative Filtering}. In
  \bibinfo{booktitle}{\emph{SIGIR}}. \bibinfo{pages}{165--174}.
\newblock


\bibitem[\protect\citeauthoryear{Wu, He, Sun, Chen, and Ye}{Wu
  et~al\mbox{.}}{2019}]%
        {HOT}
\bibfield{author}{\bibinfo{person}{Bin Wu}, \bibinfo{person}{Xiangnan He},
  \bibinfo{person}{Zhongchuan Sun}, \bibinfo{person}{Liang Chen}, {and}
  \bibinfo{person}{Yangdong Ye}.} \bibinfo{year}{2019}\natexlab{}.
\newblock \showarticletitle{ATM: An Attentive Translation Model for Next-Item
  Recommendation}.
\newblock \bibinfo{journal}{\emph{IEEE Transactions on Industrial Informatics}}
  \bibinfo{volume}{16}, \bibinfo{number}{3} (\bibinfo{year}{2019}),
  \bibinfo{pages}{1448--1459}.
\newblock


\bibitem[\protect\citeauthoryear{Yuan, He, Jiang, Guo, Xiong, Xu, and
  Xiong}{Yuan et~al\mbox{.}}{2020}]%
        {Future}
\bibfield{author}{\bibinfo{person}{Fajie Yuan}, \bibinfo{person}{Xiangnan He},
  \bibinfo{person}{Haochuan Jiang}, \bibinfo{person}{Guibing Guo},
  \bibinfo{person}{Jian Xiong}, \bibinfo{person}{Zhezhao Xu}, {and}
  \bibinfo{person}{Yilin Xiong}.} \bibinfo{year}{2020}\natexlab{}.
\newblock \showarticletitle{Future Data Helps Training: Modeling Future
  Contexts for Session-based Recommendation}. In
  \bibinfo{booktitle}{\emph{WWW}}. \bibinfo{pages}{303--313}.
\newblock


\bibitem[\protect\citeauthoryear{Yuan, Karatzoglou, Arapakis, Jose, and
  He}{Yuan et~al\mbox{.}}{2019}]%
        {NextItNet}
\bibfield{author}{\bibinfo{person}{Fajie Yuan}, \bibinfo{person}{Alexandros
  Karatzoglou}, \bibinfo{person}{Ioannis Arapakis}, \bibinfo{person}{Joemon~M
  Jose}, {and} \bibinfo{person}{Xiangnan He}.} \bibinfo{year}{2019}\natexlab{}.
\newblock \showarticletitle{A Simple Convolutional Generative Network for Next
  Item Recommendation}. In \bibinfo{booktitle}{\emph{WSDM}}.
  \bibinfo{pages}{582--590}.
\newblock


\end{thebibliography}

\end{document}